\documentclass[12pt,preprint]{aastex}

\newcommand{\ea}{et al.}
\newcommand{\myr}{\>{\rm Myr}}
\newcommand{\kms}{\>{\rm km}\,{\rm s}^{-1}}

\newcommand{\msun}{\>{M_{\odot}}}
\newcommand{\msunpyr}{\>{\msun\,{\rm yr}^{-1}}}

\newcommand{\bfi}{\begin{figure}[htb]} 
\newcommand{\bpfi}{\begin{figure}[p]}

\newcommand{\halpha}{$\rm H\alpha$}
\newcommand{\hbeta}{$\rm H\beta$}
\newcommand{\as}{ ^{\prime\prime}}
 
\newcommand{\escm}{$\rm erg\,s^{-1}cm^{-2}$}
\newcommand{\es}{$\rm erg\,s^{-1}$}

\def\deg{\hbox{$^\circ$}}
\def\hii{\relax \ifmmode {\rm H\,{\sc ii}}\else H\,{\sc ii}\fi}

\def\fdg{\hbox{$.\!\!^\circ$}}

\def\farcm{\hbox{$.\mkern-4mu^\prime$}}
\def\farcs{\hbox{$.\!\!^{\prime\prime}$}}
\def\degd#1.#2{ #1\fdg#2 }                 
\def\mind#1.#2{ #1\farcm#2 }               
\def\secd#1.#2{ #1\farcs#2 }               

\def\aj{AJ}                   
\def\araa{ARA\&A}             
\def\apj{ApJ}                 
\def\apjl{ApJ}                
\def\apjs{ApJS}               
\def\apss{Ap\&SS}             
\def\aap{A\&A}                
\def\mnras{MNRAS}             
\def\nat{Nature}              

\shorttitle{Star Formation in Nuclear Rings}
\shortauthors{	Mazzuca, et al.}

\begin{document}


\title{A Connection between Star Formation in Nuclear Rings and their Host Galaxies}


\author{Lisa M. Mazzuca}
\affil{NASA Goddard Space Flight Center, Code 441, Greenbelt, MD 20771}

\author{Johan H. Knapen}
\affil{Instituto de Astrof\'\i sica de Canarias, E-38200 La Laguna, Spain}
\affil{Centre for Astrophysics Research, University of Hertfordshire, Hatfield, Herts AL10 9AB, U.K.}

\author{Sylvain Veilleux}
\affil{Department of Astronomy, University of Maryland, College Park, MD 20742}

\author{Michael W. Regan}
\affil{Space Telescope Science Institute, 3700 San Martin Drive, Baltimore, MD 21218}


\begin{abstract}

We present results from a photometric \halpha\ survey of 22 nuclear rings, aiming to provide insight into their star formation properties, including age distribution, dynamical timescales, star formation rates, and galactic bar influence. We find a clear relationship between the position angles and ellipticities of the rings and those of their host galaxies, which indicates the rings are in the same plane as the disk and circular. We use population synthesis models to estimate ages of each \halpha-emitting (H{\sc ii}) region, which range from 1 Myr to 10 Myrs throughout the rings. We find that approximately half of the rings contain azimuthal age gradients that encompass at least 25$\%$ of the ring, although there is no apparent relationship between the presence or absence of age gradients and the morphology of the rings or their host galaxies. NGC~1343, NGC~1530, and NGC~4321 show clear  bipolar age gradients, where the youngest \hii\ regions are located near the two contact points of the bar and ring. We speculate in these cases that the gradients are related to an increased mass inflow rate and/or an overall higher gas density in the ring, which would allow for massive star formation to occur on short timescales, after which the galactic rotation would transport the \hii\ regions around the ring as they age. Two-thirds of the barred galaxies show correlation between the locations of the youngest \hii\ region(s) in the ring and the location of the contact points, which is consistent with predictions from numerical modeling.

\end{abstract}


\section{Introduction}

Nuclear rings are viewed as a type of starburst, consisting of distinct compact areas of young massive stars near the centers of spiral galaxies.  The intense bursts, attributed by a Star Formation Rate (SFR) that significantly exceeds the average found among normal galaxies, can dominate the overall star formation in early Hubble type barred spiral galaxies, and dramatically alter the structure of the host galaxy with respect to morphology, gas and dust content, metallicity, and stellar population (e.g., Martinet 1995; Buta \& Combes 1996; Kormendy \& Kennicutt 2004). Typically located within the central kiloparsec, nuclear rings contain a mixture of neutral and ionized gas and dust with masses of $10^{8}$ to $10^{10} \msun$ (Heller \& Shlosman 1996; Rubin, Kenney, \& Young 1997). 

The great majority of nuclear rings are found in barred spiral galaxies, where the bar provides essential fuel for the formation and evolution of the ring via shock-induced channeling of disk gas to the nuclear regions (e.g., Schwarz 1981; Combes \& Gerin 1985; Shlosman, Begelman, \& Frank 1990;  Athanassoula 1992; Knapen \ea\ 1995b; Heller \& Shlosman 1994; Heller \& Shlosman 1996). The shocks can act as instigators for angular momentum loss, which is crucial for directing the gas towards the central regions and forming nuclear rings.  In such a picture massive SF would result and be most intense near where the bar dust lanes intersect the ring, called the contact points. The contact points can lead the bar by as much as 90\deg\ with respect to the galactic rotation (Knapen \ea\ 1995b; Heller \& Shlosman, 1996; Regan, Vogel, \& Teuben 1997; Regan \ea\ 1999; Ryder, Knapen, \& Takamiya 2001; Allard, Peletier, \& Knapen 2005).
  
While many galaxies with nuclear star-forming rings have been identified (Knapen 2005 states that 20\% of local spirals host them), it is hard to generalize on the physical mechanisms behind the triggering and propagation of massive SF due to the inconsistent and incomplete nature of the existing data. To allow for a more statistical approach to answering some of the mysteries of these rings, we performed an imaging survey of 73 galaxies in \halpha\ and the $B$ and $I$ bands, many of which host nuclear rings. An overview of this survey is presented in Knapen et al. (2006; hereafter Paper~I). In Paper~I we classified the morphology of the nuclear and circumnuclear H{\sc ii} regions and confirmed that (1) most late-type galaxies have a patchy circumnuclear appearance in \halpha, (2) nuclear rings occur primarily in spiral types Sa-Sbc, and (3) found that the presence or absence of a close companion does not significantly affect the \halpha\ morphology around the nucleus nor the strength of the nuclear \halpha\ peak.

Of these 73 galaxies, we identified 22 that host nuclear rings as judged from our \halpha\ images, of which 19 were previously known as such.  This paper isolates the individual \hii\ regions forming the nuclear rings. We present information on the observations and data reduction in Section~2. We derive morphological parameters of the rings using ellipse fitting, and then compare those to the disk characteristics, in Section~3. In Section~4, we identify and analyze the \hii\ regions forming the rings, compute their equivalent widths (EWs), and estimate error budgets. Using evolutionary models, we convert the EWs of the \hii\ regions to ages and discuss the assumptions we make with respect to the type of starburst activity occurring in and around the rings in Section~5. We present our age distribution analysis, star formation rate estimations, and a discussion of the stellar bar dynamics with respect to the rings in Section~6, before summarizing our conclusions in Section~7. 

\section{Observations and Data Reduction}\label{obs} 

The ring images presented here are part of a larger William Herschel Telescope (WHT) imaging survey conducted to characterize the nuclear and circumnuclear morphology in nearby galaxies (Paper~I). We selected galaxies with some evidence for \halpha\ structure in their central regions, either from the literature (e.g., Buta \& Combes 1996; Buta \& Crocker 1993) or from our own past work. We adopted the definition of a nuclear ring, nuclear spiral, and psuedo-ring from Buta \& Crocker (1993) in Paper~I and follow it here for consistency.

Since one of the aims of our overall study is the detailed analysis of nuclear rings, our sample was biased towards SBC galaxies known to host such features. We chose rings with enough structure to analyze, and which are considered to be the brightest feature in the circumnuclear region. Given the nature of our original survey from which the targets were selected, a distance bias is to some extent unavoidable.  We recognize the wide range of distances and ring sizes in the sample, and confirm that  rings in farther galaxies are larger than those in nearby galaxies. However, since our selection was with respect to the angular size of the ring, we find that there is no correlation of distance to the number of identifiable \hii\ regions. 

The images presented in this paper have been obtained using the Auxiliary Port (Aux Port) camera on the 4.2m WHT, operated by the Isaac Newton Group in La Palma. The camera's circular field of view (FOV) is $\sim$1$\farcm$8 in diameter, imaged with a 1024 $\times$ 1024 pixel TEK CCD. The angular sizes of the nuclear rings in our sample vary between $5\as$ and $30\as$, which fit well into the FOV. We used Harris $B$ and $I$ filters, and one of four narrow-band \halpha\ filters, depending on the systemic velocity of the galaxy. We refer the reader to Table~1 of Paper~I for a listing of the complete sample, along with their global parameters. Table~1 of the present paper lists morphological type, ring ellipticity, position angle (PA) of the disk and bar, and derived ring size for our subsample of 22 galactic nuclear rings as identified in Paper~I.

We observed spectrophotometric standards to calibrate the \halpha\ images, and photometric standards to calibrate the {\it B} and {\it I} band images. We observed the standards through the same filters as the galaxies, with exposure times ranging from 5 to 20 seconds. The instrumental magnitude of each star in each filter was obtained using the IRAF package {\sc phot}. This photometry package calculates the magnitude, among other parameters, based on user-defined values for the radius and annulus of a sky aperture. To convert from instrumental magnitude to AB magnitude for the spectrophotometric standard stars, an interpolation method was used based on results from Oke (1990). Oke defines values of AB at points where no ambiguity exists for instrument calibration problems due to spectral resolution. Therefore, to obtain magnitudes for the filters used in our survey, magnitudes with respect to the wavelengths just above and below the interested filter wavelength were chosen. Interpolating between these values yields the absolute magnitude for the five \halpha\ filters. The resulting integrated fluxes of the rings (on the order of 10$^{-12} - 10^{-14}$ \escm) and luminosities of the \hii\ regions (on the order of 10$^{38} - 10^{40}$ \es; see Appendix A) are consistent with published data (e.g., Pogge 1989; Storchi-Bergmann, Wilson, \& Baldwin 1996; Regan \ea\ 1996; Sheth \ea\ 2000).

\section{Ring Morphology}
We fitted the nuclear rings of the 22 galaxies with the IRAF task {\sc ellipse}. We obtained the central position of each ring from the peak in the $I$-band image, and fixed the center while iteratively fitting elliptical isophotes to the \halpha\ image. We masked out the nucleus and foreground stars that could affect the fitting process, and provided initial estimates of the major axis length, ellipticity, and PA. The task produced measurements of the azimuthally integrated intensity, observed ellipticity, and PA of the ring major axis as a function of radius. We determined the best-fitting ellipse, which corresponded to the peak isophotal intensity across the radial range of the ring (see Table~1). Figure~1 shows the resulting fits of the 22 rings, overlaid on the \halpha\ images. The pixel-based ring diameters produced from {\sc ellipse} were converted to arcseconds and kiloparsecs (see Table~1, Columns 5 \& 6, respectively). The rings vary in major axis from 0.2~kpc to 1.7~kpc, which is typical for nuclear rings (Buta \& Crocker 1993). In most cases, the galaxy distance was obtained from the Nearby Galaxies Catalog (Tully 1988); in the cases where there were no catalog values recorded, we computed them by dividing the recessional velocity by the Hubble constant (taken at 75\,${\rm km}\,{\rm s}^{-1}\,{\rm Mpc}^{-1}$) -- see Paper~I for details. Table~1 lists the parameters of the isophote best representing each ring, chosen on the basis of the peak isophotal intensity across the radial range of the ring.

We compare the ellipticity and PA of each ring to that of its host disk. Disk ellipticities and PAs are from the RC3, with the exception of the values for NGC~1530, which are based on the kinematical data of Regan \ea\ (1996). Figure~2 compares the PAs and ellipticities of the rings to their host disks. The plots show a clear linear relationship between the ring and its host disk with respect to both PA and ellipticity. This indicates that the rings are in the same plane as the disk and are nearly circular when deprojected. In some cases (NGC~278, NGC~4303, NGC~4314, IC~1438, and NGC~7742) the disk is nearly circular; those galaxies are seen face-on and do not appear in Figure~2. The disk morphology in these cases is consistent with that of the rings, which all have very low ellipticities (less than 0.1). The only exception to this general statement is NGC~7570 whose ring PA and ellipticity differ significantly from those of the disk. The ring appears to be nearly face-on so the PA of the chosen isophotal fit may not be reliable. However, the disk high inclination ($i$=54\deg; RC3) is not consistent with the observed nearly circular ring. This strongly suggests that the ring is not in the same plane as the disk.

\section{Equivalent Widths of H{\sc ii} Regions in Nuclear Rings}

\subsection{Identification of H{\sc ii} Regions}
To identify the individual \hii\ regions forming a given ring, we used the program {\sc SExtractor} (Bertin \& Arnouts 1996), which builds a catalog of \hii\ regions and their associated fluxes from an astronomical image.  The various parameters used allow for the detection of \hii\ regions even at very low \halpha\ intensities. We masked out the nucleus and extraneous stars that were clearly not part of the ring. The catalog file produced by the program records the position, area, and flux of each \hii\ region detected. Figure~3 shows the outlines of the fitted \hii\ regions overlaid on the continuum subtracted \halpha\ images. In the case of NGC~7469, {\sc SExtractor} was unable to identify any \hii\ regions due to poor spatial resolution compared to the small angular size of the ring. This galaxy is well known to contain a nuclear ring of 0$\farcs$5 in radius (Genzel \ea\ 1995; Davies, Tacconi, \& Genzel 2004; D\'\i az-Santos \ea 2007), which is consistent with our {\sc ellipse} results (Section~3). The lack of sufficient spatial resolution in our images, however, prevents us from continuing with the analysis of the individual \hii\ regions in the ring, but we refer to above citations for a detailed study of its SF properties, based on high-resolution imaging and near-IR spectroscopy.

To obtain EWs of the \hii\ regions, we measure fluxes in both the $I$-band and \halpha. We do this by first imposing the position and shape of the apertures from the {\sc SExtractor} analysis of the \halpha\ image onto the $I$-band image, making use of the fact that these images have already been registered (see Paper~I). We then run {\sc SExtractor} in its two-image mode to obtain the corresponding $I$-band fluxes. Among the many parameters to be defined in {\sc SExtractor}, two are of particular importance, namely the initial box size for the \hii\ region detections, and the threshold value used to determine the extent of an individual \hii\ region. For the first parameter we chose a value corresponding to the average seeing value in our images, namely 8 pixels or 0$\farcs$9. For the threshold value, we chose twice the standard deviation, $\sigma$, of the combined galaxy and sky background across the whole image.


We confirmed our parameter settings by independently varying the threshold value and box size and comparing the resulting ring morphology and EWs to our "reference" values, which were best representative of the morphology seen in the \halpha\ images. Several test cases were run by altering each parameter (fixing one parameter while modifying the other) by $\pm$50$\%$ of the reference values. Figure~4 shows the resulting images for NGC~1343, as a representative example. Figures 4a and 4b are the result of fixing the threshold and altering the box size. Modifying the box size by $\pm$50$\%$ of the reference value (8$\times$8 pixel box) to a 12$\times$12 box and 4$\times$4 box, respectively, causes a significant change in the visual appearance of the \hii\ regions. Increasing the box size results in an obvious over-blending of the \hii\ regions, with the opposite occurring when decreasing the box size. Figures 4c and 4d of Figure~4 are the products of modifying the threshold by $\pm$50$\%$ of the reference value (2$\sigma$), to 1.33$\sigma$ and 3$\sigma$, respectively.  Changing the threshold in either direction causes a significant change in the number of \hii\ region detections forming the ring. Overall, the isophotal detection area is more sensitive to the box size, while the number of detections is more sensitive to the threshold. For small variations of the threshold, the majority of the detected regions remained in tact. Added detections were either as a result of further segmenting an already-defined region or adding a new region that was now above the noise level. These parameters are important since the background to be subtracted from the $I$-band and \halpha\ data is locally defined by {\sc SExtractor} for each \hii\ region, and includes both the sky background and the emission from the galactic disk. In general, this empirical analysis gives us confidence that our parameter choices do indeed best represent the \hii\ region morphologies.
 
\subsection{Equivalent Widths}
To obtain EWs for the individual \hii\ regions, the \halpha\ flux was divided by the {\it I}-band flux within the same aperture as defined by {\sc SExtractor}. We describe below some of the intricacies of the EW determination. We specifically discuss the correction of those cases where the $I$-band flux of an \hii\ region is particularly low, as well as the error budget of the EW values. 

\subsubsection{Negative $I$-band Fluxes}
Many of the galaxies in our sample (16 out of 20) contained \hii\ regions whose $I$-band emission was lower than the background due to noise. This resulted in negative $I$-band fluxes, which are unphysical and cannot be used to determine the EW. Rather than changing either the \hii\ region aperture or specific {\sc SExtractor} parameters, we chose to use a physically meaningful upper limit to the $I$-band flux. We did this by taking, for all galaxies, the lowest $I$-band flux per pixel ($f_{\rm min}$) of all of the \hii\ regions. Analyzing the values per pixel versus per region avoids the possible bias introduced by the large range of region sizes, which vary by as much as a factor of two in a given ring. We conservatively assume the lowest value to be representative of a 6$\sigma$ deviation from the mean. Where $f_{\rm min}$ is negative, we correct any $I$-band flux per pixel lower than the value $f_{\rm corr}= \frac{1}{2}|f_{\rm min}|$ (equivalent to a 3$\sigma$ adjustment) to have a corrected \hii\ region flux $F_{\rm corr} = f_{\rm corr}\times{\rm area}$. This method, thus, applies a correction to those \hii\ regions whose emission is positive but very low. See Appendix~A for details on individual \hii\ region corrections.

To estimate the impact of the 3$\sigma$ corrections on the EWs, we also corrected these fluxes to a lower level, namely $\frac{1}{3}|f_{\rm min}|$ (equivalent to a 2$\sigma$ adjustment). The maximum impact on EW that we found in such cases was an increase in EW of order 0.2 in log(EW), although in many cases the difference was less than 0.1. In 10 out of the 16 galaxies affected by negative $I$-band emission, 5\%, on average, of the \hii\ regions subject to the 3$\sigma$ correction would not be subject to a 2$\sigma$ correction. That is, those \hii\ regions with very low positive $I$-band emission that were changed using the 3$\sigma$ discriminator were no longer affected when compared to the 2$\sigma$ value. The other six galaxies were entirely unaffected by the correction change and can therefore be considered more reliable. We present these findings in Figure~5 using a 3$\sigma$ correction. Regions whose $I$-band flux was changed to the 3$\sigma$ value but was unaffected by the 2$\sigma$ comparison are labeled as a diamond with a cross; these represent lower-limits in the log(EW) since we find that these values would have increased if we had chosen a 2$\sigma$ correction. All other \hii\ regions with corrected $I$-band fluxes, including the ones that were negative, are labeled as open diamonds.  Those \hii\ regions where no corrections were needed are labeled as filled diamonds, and we consider those the most reliable.

\subsubsection{Characterizing the Uncertainties}

We estimate typical uncertainties in the EW values by considering uncertainties due to background subtraction and \hii\ region detection, which both result from the application of {\sc SExtractor}. Additional contributions to the overall error budget will come from measurement errors in the standard stars, the calibration uncertainties associated with our method of continuum subtraction (see Paper~I), and in the filter response curves, but we will assume that they are negligible compared to the two main contributors.

To make a conservative estimate of the uncertainty in the background
subtraction we consider the $I$-band backgrounds. These will be more
affected than the H$\alpha$ measurements because (1) the background is higher
in $I$ than in H$\alpha$, (2) the contrast between the \hii\ region and
background is much smaller than in H$\alpha$, and (3) the basic
morphology of an \hii\ region has been set by its H$\alpha$ appearance in the $I$-band. From our
analysis of the negative residual $I$-band fluxes, we know that
the maximum difference in the log of the EW between the two cases (i.e.,
applying a 2$\sigma$ versus 3$\sigma$ correction to the affected measurements) is
0.2. Such a difference of log(EW)=0.2 corresponds to a factor of 1.5
difference in the $I$-background subtracted before measuring the flux of an
\hii\ region, where log(EW)=0.1 would correspond to a factor of 1.3. Even
considering the small field of view of our $I$-band images, these
factors cover the uncertainty in the determination of the background.


The other main contributor of the EW uncertainties is in the flux measurement of the \hii\ regions. An RMS value, based on Gaussian noise statistics, is associated with each detection. On average, we estimate the RMS to be 5\% of the detection, which equates to 0.02 in the log of the EW. This constant RMS value is due to the fact that both flux and source size play a role in the noise (e.g., detections with weaker fluxes cover smaller areas).


We thus arrive at a conservative error budget consisting of two
components, 0.2 and 0.02 in the log(EW). These combine in a worst case
scenario to a factor of 1.6, or an error budget in log(EW) of 0.2. We take 3$\sigma$ of the total error budget as a guideline for discriminating between EWs (and thus \hii\ regions). As a consequence, two EW values must be 0.6 apart in the log to make them significantly different. Whereas this is by all means a large uncertainty, Figure~5 shows that the values can range by as much as 3 in the log for a given ring during the evolution of massive stars so that, in fact, most of our measurements lead to straightforward and believable conclusions on at least relative ages, as discussed in Section~6.1. We stress that the error budget determined here is a very conservative one, and that for most \hii\ regions the actual uncertainty will typically be significantly smaller.

%

\section{Age Dating}
We use Starburst99 (Leitherer et al. 1999) stellar population modeling to estimate the age of the individual \hii\ regions. The models present the time evolution of the EW(\halpha) for varying IMF properties and metallicities, which will affect the evolution of the population. Figure~6 shows the time evolution of EW(\halpha) for both an instantaneous starburst (i.e., all stars are formed simultaneously in an instantaneous starburst episode) and a continuous starburst (i.e., the star formation rate is constant in time). While the  starburst choice can greatly affect the determination of the ages of the population greater than $\sim$5$\myr$, it is not clear which model best represents the stellar activity in nuclear rings. Since active massive star formation is prevalent in \hii\ regions, adopting an instantaneous burst would seem to fit the model predictions of younger ages better over the range of 1$\myr$ - 10$\myr$. If, on the other hand, there is an abundant supply of gas entering the ring, starbursts may be a composite of episodic bursts (Elmegreen 1997; Allard \ea\ 2006; Sarzi \ea\ 2007). Star formation in the ring would cyclically enter an active starburst period until it exhausts the surrounding gas supply, and then cease star formation until enough gas sparks a new starburst period. 

Figure~7 shows Starburst99 model curves for an instantaneous and a continuous starburst with different IMF slopes, mass cut-offs, and metallicities. We vary the IMFs in both scenarios and find that the evolution model is not very sensitive to the upper limit masses (i.e., 30$\msun$ and 100$\msun$) or the IMF slopes ($\alpha = 2.35, 3.3$), especially in the instantaneous starburst case. The models also confirm that high metallicity \hii\ regions show lower EW(\halpha) while low metallicity regions have invariably high EW(\halpha). Regardless of the metallicity or type of burst, however, we find that all of the predicted evolution curves show a significant monotonic drop of EW with increasing age. Moreover, the variation shown in the model curves (Fig.~7) is most likely an overestimate of the real variations in the
nuclear rings under consideration here. For instance, Allard et al. (2006) and Sarzi et al. (2007) found Solar metallicity values for all nuclear rings in their sample, and there is no convincing evidence for changes in either the IMF slope or upper mass cutoff, except possibly in rather extreme environments such as the Galactic Center (e.g., reviews from Elmegreen 2002 and Kroupa 2002). Since we are more interested in the relative age difference from one \hii\ region to the next, the sharply declining curve (in either starburst case) allows for an unambiguous discussions of age gradients. We adopted an instantaneous stellar starburst with a Salpeter (1955) IMF between $M_{\rm low} = 1\msun$ and $M_{\rm up} = 100\msun$, and a solar metallicity, which is a good predictor for circumnuclear regions of barred galaxies (Allard \ea\ 2006).

\section{Discussion}
\subsection{Age Distribution}
The results from {\sc SExtractor} were used to map the ages of each \hii\ region according to their azimuthal location in the ring. The right panels of Figure~5 show the resolved \hii\ regions, which are color-coded according to age, in terms of EW. We assigned the physical center of the ring (as computed with {\sc ellipse}) to be (0,0), and viewed an \hii\ region's azimuthal position in a counter-clockwise rotation from North to East. The color ranges are unique to each ring and, thus, do not denote the same EWs for all rings. Although we did not deproject the rings, all \hii\ regions have a unique position, which allows us to search for any unambiguous distribution patterns. The PA of the major axes of the ring and the bar are indicated in the image. To complement the images, plots are shown in the left panels, which graph the log(EW) versus PA, and again indicate the PAs of the ring and bar major axes. 

We use $B-I$ color index maps, shown in Figure~8, as morphological indicators of the ring kinematics. These maps provide direct tracers of the bar dust lanes, which show the flow of gas to the ring. In the cases where the dust lanes were present, the azimuthal direction of the ring rotation was consistent with that of the galaxy in all cases. The contact points of the dust lanes with the ring are clearly visible in most of our images and indicate an offset of $\sim$90$\deg$ between the contact points and the bar (see Section~6.3 for details). The dust lanes also show directional flow around the ring, assuming that rotation continues in the same direction as seen with the bar dust lanes and spiral arms. The adopted rotational direction of the rings is listed in Table~2 and indicated for each ring containing a bar in Figure~5.

We find that several galaxies, both barred and non-barred, exhibit relative azimuthal age gradients within the ring, as also observed by Ryder, Knapen, \& Takamiya (2001) and Allard \ea\ (2006) for NGC~4321. We only consider age gradients over a significant portion of the ring (greater than 25\% of the ring) with amplitudes that exceed the uncertainties for differentiating between points over the range of interest. We highlight our findings below, and summarize the characterization of the age distributions for all of the rings (e.g., azimuthal gradient, radial gradient, flat distribution, or no recognizable pattern) in Table~3 (Column~2).

\begin{itemize}
\item NGC~278 (Fig.~5a):  The entire ring as seen in \halpha\ is very small in size (0.2 kpc in radius) with 20 individual \hii\ regions detected that comprise the ring. This ring is the smallest one recognized in the galaxy; a second ``outer'' nuclear ring appears at $\sim$ 1~kpc and is discussed in detail in Knapen \ea\ (2004). This outer ring is considered 'non-standard' as it is very wide, and more a region of star formation with rather abrupt delimitations on the inner and outer radial range. In addition to this, we were in practice unable to derive a recognizable ring pattern of \hii\ regions using {\sc SExtractor}, and have thus opted for concentrating on the inner nuclear ring.  The host galaxy does not contain a bar in the optical nor near-IR wavelengths (Knapen \ea\ 2004). There is a decrease in the log(EW) (i.e., an increase in age) over approximately one third of the inner ring ranging from the positions of the \hii\ regions centered at azimuth 124$^\circ$ to 236$^\circ$. This indicates a counterclockwise rotation within the ring. A radial gradient spans the same range, with the younger \hii\ regions lying on the inner side of the ring. A similar pattern, although not as clearly defined, appears over the range of 12$^\circ$ to 100$^\circ$.

\item NGC~473 (Fig.~5b):  Two-thirds of the ring can be seen in the \halpha\ image, defined by 27 \hii\ regions. A galactic bar exists at a major axis PA of 164$^\circ$ along the major axis. The bar dust lanes indicate a counterclockwise flow of material around the ring. There is a sharply sloped age gradient over the range of 38$^\circ$ to 203$^\circ$ from younger to older, respectively.  

\item NGC~613 (Fig.~5c):  There exists a small partial nuclear ring containing 11 individual star-forming \hii\ regions with a strong nucleus. A fairly flat distribution in the log(EW) appears around the entire ring with ages averaging no less than 10 Myrs. The galactic bar is located at PA~=~110$^\circ$ along its major axis. Although a gradient is not defined, one of the contact points (at 200$^\circ$) coincides within 5$^\circ$ to the youngest hotspot (PA~=~205$^\circ$) in its respective hemisphere.

\item NGC~1300 (Fig.~5d):  Twelve \hii\ regions form a small complete ring in \halpha\, although there is no clear pattern in EWs. 

\item NGC~1343 (Fig.~5e):  The ring in this barred spiral galaxy is seen as complete in \halpha\ with 20 resolved \hii\ regions. The mapping of EW to \hii\ region location indicates a clear bi-polar age pattern around the ring with EW maxima (i.e., relative youngest age in the ring) at PA~=~140 and 309$^\circ$. There is a distinct age gradient approximately from one contact point to the other contact point (located at 172$^\circ$ and 352$^\circ$, respectively), where the age increases counterclockwise from the bar contact points.

\item NGC~1530 (Fig.~5f):  Three-quarters of the ring can be seen in \halpha\, containing 9 distinct \hii\ regions. The galaxy has a large bar at PA~=~122$^\circ$ and two open spiral arms originating from its ends, rotating clockwise. We find bipolar age gradients, one in each hemisphere, as bisected by the position of the bar. The youngest \hii\ regions are near the bar PA, (i.e., not the contact points) over the range of 283\deg\ to 207\deg\ and 115\deg\ to 7\deg, with the ring rotation also clockwise. Half of the ring is distinctly younger than the other half, where the turnover points coincide with the ring major axis PA (= 25$^\circ$). 

\item NGC~4303 (Fig.~5g):  Only seven \hii\ regions define the small ring. The ring is observed to be three-quarters complete in \halpha\ with bipolar age maxima at PA~=~114$^\circ$ and 304$^\circ$. There are no obvious age gradients in between the maxima, but the latter are offset by less than 25$^\circ$ from the bar contact points (100$^\circ$ and 280$^\circ$).

\item NGC~4314 (Fig.~5h):  The small ring consists of 10 \hii\ regions showing a smooth and well-defined age gradient. The age gradient flows from older to younger in the same direction as the presumed ring rotation (clockwise). The youngest \hii\ region is within 20$^\circ$ of one of the contact points (PA~=~45$^\circ$).

\item NGC~4321 (Fig.~5i):  The well resolved full ring contains 57 detected \hii\ regions. Because missing data prevent the calculation of \halpha\ equivalent widths and ages, we refer to Allard \ea\ (2006) for a similar analysis in \hbeta. They find a bipolar age gradient corresponding to the ring PA (at 170\deg\ and 350\deg) and the location of the youngest \hii\ region in each hemisphere. The age increases in a counter-clockwise direction along the direction of flow in the ring. 

\item NGC~5248 (Fig.~5j):  Nineteen \hii\ regions define the full, elliptically shaped ring. The youngest \hii\ region is located 4$^\circ$ away from of one of the bar contact points (PA~=~47$^\circ$).

\item NGC~5728 (Fig.~5k):  Eighteen \hii\ regions define the partial ring, which is three-quarters complete. A very steep but small gradient exists from PA~=~147$^\circ$ to PA~=~98$^\circ$ from younger to older, according to the clockwise directional flow of the ring. One contact point (PA~=~303$^\circ$) is within 2$^\circ$ of the youngest \hii\ region in the ring. The second youngest \hii\ region (PA~=~147$^\circ$) is within 24$^\circ$ of the second contact point (PA~=~123$^\circ$), although it does precede that contact point. 

\item  NGC~5905 (Fig.~5l): The ring is small but complete in \halpha. It consists of nine \hii\ regions and is relatively flat with respect to age (except for one \hii\ region). There appears to be no correlation in age with respect to the bar contact points.

\item NGC~5945 (Fig.~5m):  Only three \hii\ regions define this patchy ring, which contains low EW values (all less than 1.5 in the log). Buta \& Crocker (1993) indicate a nuclear ring, but our images do not have enough resolution to confirm this.

\item NGC~5953 (Fig.~5n):  The non-barred galaxy has a nuclear ring that appears complete with 23 \hii\ regions. Although we know of no bar to directly fuel the ring, it is well formed and resolved both azimuthally and radially.

\item NGC~6503 (Fig.~5o):  The elliptically shaped ring contains 84 resolved \hii\ regions in this unbarred spiral. No gradients are evident.

\item NGC~6951 (Fig.~5p):  The ring is small but complete with eight detected \hii\ regions. The youngest \hii\ region (PA~=~339$^\circ$) is within 16$^\circ$ of its respective contact point. The ages steeply increase (by a factor of 10) from 339$^\circ$ to 218$^\circ$ in a clockwise direction, consistent with the directional flow of the galaxy.

\item NGC~7217 (Fig.~5q): The ring is three-quarters complete in \halpha. Twenty-two \hii\ regions were detected in the ring of this unbarred galaxy. Ages increase in a clockwise direction from  PA~=~109$^\circ$ to PA~=~6$^\circ$ .

\item IC~1438 (Fig.~5r):  This barred galaxy hosts a complete ring with 11 \hii\ regions detected. There exists an increasing age gradient from PA~=~32$^\circ$ to PA~=~211$^\circ$ in a counterclockwise direction. This direction is consistent with the ring rotational flow. The youngest \hii\ region is within 2$^\circ$ of its respective bar contact point (PA~=~304$^\circ$)

\item NGC~7570 (Fig.~5s):  Eighteen \hii\ regions comprise a mostly complete ring in this barred galaxy. An increasing age gradient ranges counterclockwise from PA~=~112$^\circ$ to  PA~=~295$^\circ$. There is also a radial gradient spanning the same range as the azimuthal gradient, with two distinct arcs at nearly identical slopes, and where the inner arc appears consistently older than the outer one. No correlation to the bar contact points is evident. 

\item NGC~7716 (Fig.~5t):  Thirty \hii\ regions form a complete ring in this barred galaxy. No gradients or correlations to the bar are observed.

\item NGC~7742 (Fig.~5u):  The circular and well resolved ring contains 38 \hii\ regions. No gradients are observed.

\end{itemize}

While many of the rings contain age gradients, a few show no gradient, namely NGC~613, NGC~5905, and NGC~5945. This type of distribution is qualitatively more consistent with Elmegreen's (1994) predictions, which suggest that the massive star formation in the nuclear ring is shocked into existence due to gravitational instabilities in the nuclear region around the ILRs. Therefore, age gradients would not exist. Comparisons of EW to age for these three rings also indicate that these starbursts are relatively older than those seen in the other rings, on the order of 10 Myrs and higher. Elmegreen (1997) suggests that these rings (i.e., with no gradient) could last for as long as 100 Myrs before SF ceases if the inflowing gas rate is high enough. The SFRs of these three rings (2$\msun$/yr, 2$\msun$/yr, and 4$\msun$/yr, respectively; see Section~6.2 and Table~1 for SFR and inflow discussion) are among the higher values in our sample but are consistent with other rates documented (Regan \ea\ 1997; Jogee \ea\ 2002). If we take into account that NGC~613 and NGC~5945 are partial rings in the optical, then the SFRs calculated here are lower limits.

\subsection{Inflow and  Star Formation Rates}

In this Section we explore possible reasons for the result that only three galaxies show bi-polar age gradients around their rings, whereas many of the rings (9 of 22) show partial gradients along 25\% - 30\% of their rings, with the rest (10 of 22) showing no gradient. 

We find no apparent relationships between the presence or absence of age gradients and the morphology of the rings or their host galaxies. For example, whereas NGC~4321, with its bi-polar age gradient, has a pair of well-defined dust lanes coming into the ring through the bar, NGC~1300 and NGC~5248 have similar dust lane morphologies yet show no gradients. Some of the poorly resolved rings, such as those in NGC~4303 or NGC~5945, do not show gradients, but given the presence of measured age gradients in the comparably small rings in NGC~6951 and IC~1438, the lack of spatial resolution in the former two cannot easily be used as an argument for the lack of a gradient. But even if we ignore galaxies with either small and badly resolved rings, or rings that do not have the classical aspect of dynamically-induced resonance rings (NGC~613 and NGC~6503 are in the latter category), we are left with a significant number of rings in which we should have detected gradients if present, but which for some reason do not show any. This category includes, for instance, NGC~1300, very similar to NGC~4314 or NGC~6951 which do show gradients (all compact well-defined rings with around a dozen \hii\ regions, sitting in strong bars with well-defined sets of dust lanes); NGC~5248, similar to NGC~4321 (complex, rich rings with many \hii\ regions organized into spiral arms fragments and hosted by a relatively weak bar); NGC~7716, similar to NGC~7217 which does have a gradient; and NGC~7742. For those rings that contain a gradient we also find no correlation between the presence of the gradient and direction of the ring rotation.

In relating the location of the youngest \hii\ regions to those of the contact points between the dust lanes and ring, we have been assuming that massive SF will follow very rapidly once the inflowing gas joins the ring.  The appearance of a gradient in the ring may thus depend on whether or not material flowing in will immediately trigger gravitational collapse, and initiate massive SF. This in turn may be due to either a higher mass inflow rate (or possibly increased clumpiness of the inflow) and/or an overall higher existing gas density in the ring. Each of these possibilities will ensure that soon after a clump of inflowing gas joins the ring, it will start to form massive stars. The alternatives, of low inflow rate and/or overall low gas density in the ring, might lead to a more gradual increase of the gas density, without necessarily leading to immediate gravitational collapse and massive SF on short timescales. In this case, the positions of the \hii\ regions would bear no relation to the locations of the contact points. The ring would, in such a case, merely increase its gas density gradually, until at some stage, somewhere along the ring, an individual clump would become unstable to collapse; therefore, no age gradients would ensue.

To test this simple model, we must first consider the quantities mentioned above, gas inflow rate and gas density in the ring. These can in principle be determined from interferometric CO measurements, but the process is not straightforward. The main issues are the conversion factor of $L_{\rm CO}$ to $M_{\rm H_2}$ (the $X$-factor), which may well not be equal to the canonical value in the kind of environments we consider here, and the determination of the fractions of detected molecular gas which are in fact flowing into the ring rather than around it (as a consequence of a ``spray-type'' flow, see, e.g., Regan, Vogel \& Teuben 1997; Jogee et al. 2002), or with the right physical conditions such as velocity dispersion and clumpiness to take part in the massive SF process. So although CO interferometry has been presented for seven of our nuclear rings in the literature (by Regan, Vogel \& Teuben 1997; Sakamoto et al. 1999; Jogee et al. 2002, 2005; Helfer et al. 2003; Combes et al. 2004; Garc\'\i a-Burillo et al. 2005), we will not attempt here to derive the gas inflow rate and gas density in the ring from this collection of data, all with different observational characteristics. For two of our rings, NGC~1530 and NGC~5248, gas inflow rates have been explicitly determined in the literature. For the former galaxy, Regan et al. (1997) give a value of $\dot{M}=1\msunpyr$, whereas Jogee et al. (2002) determine a value of ``a few''$\msunpyr$ for the latter galaxy. Considering the uncertainties in these values, we can say little more than that they are consistent with the SFRs determined for our nuclear rings (see discussion below, and values in Table~1).

We could assume that the SFR of the ring can act as a tracer of the inflow rate, although that would imply assuming, for instance, that inflowing gas is transformed into massive stars on very short timescales. If we do this, we would expect the age gradients to occur in those galaxies with the higher SFRs, which would be representative of higher gas inflow rates. In contrast, lower SFRs would indicate a low inflow rate so that the gas will gradually fill the ring, leading to SF only if and when the local gas density leads to gravitational collapse. This will occur essentially at random and thus result in a ring with no gradients.

Kennicutt (1998) derives a relation between the \halpha\ ionizing flux and the SFR for normal disk star forming galaxies, where the conversion factor is computed using stellar population synthesis modeling. Since only stars with masses larger than 10\,$\msun$ and lifetimes less than 20\,$\myr$ dominate the integrated ionizing flux, the \halpha\ emission line provides a nearly instantaneous and linear measure of the SFR, independent of the previous star formation history. For solar abundances and our adopted IMF ($\alpha = 2.35$; $M_{\rm low} = 1\msun$ and $M_{\rm up} = 100\msun$), we can use Kennicutt's (1998) equation:

$${\rm SFR}(M_\odot\,{\rm yr}^{-1}) = 
7.9\times10^{-42}\,L({\rm H}\alpha)\,(\rm erg\,s^{-1}).$$
to compute SFRs integrated over the entire ring, which are given in column~10 of Table~1. 

Variable extinction certainly adds uncertainty in the computation of the SFR for circumnuclear regions. From the $B~-~I$ images, the patchiness of the extinction is very clear in several of our galaxies (e.g., NGC~473, NGC~613, NGC~1530, NGC~5945), which reveal only partial rings in \halpha. The less obscured regions will contribute more heavily to the line fluxes in these cases. The $B~-~I$ images clearly indicate that while the dust lanes can be traced nicely in the bar inmost cases (where  there is little star formation), the appearance of the nuclear rings is dominated by blue star-forming regions. This type of morphology makes it difficult to precisely model the effect of the dust, but attempts have been made for nuclear regions in particular. Osmer, Smith, \& Weedman (1974) analyzed six galaxies and derived integrated extinction values across the entire circumnuclear region ranging from $A_{{\rm H}\alpha}\approx2.5-3$~mag, while Hummel, van der Hulst \& Keel (1987) found an extinction of $A_{{\rm H}\alpha}\approx1.1$ in the central region of NGC~1097. Kennicutt (1998) acknowledges the high uncertainty and estimates that the errors in the SFR can be as high as $\pm$50\%. A high-resolution survey paper by Phillips (1996) for nine circumnuclear regions confirms the potential for high uncertainty values. Following Kennicutt, he finds that the SFRs vary by as much as a factor of ten when applying extinctions of $A_{{\rm H}\alpha}$=1.1 and $A_{{\rm H}\alpha}$=3.4~mag. He concludes that circumnuclear SF can be a dominant factor to the overall rates of such galaxies. We thus place the same caution that our SFRs contain potentially high uncertainties due to variable extinction throughout the nuclear rings.

Although there are possible large uncertainties in the SFR due to extinction, using the SFR can provide a possible link to an explanation for the existence of age gradients, via the gas inflow rate.   Figure~9 plots the integrated SFR over the entire ring for all galaxies and distinguishes those rings that show a gradient and those that do not. There is no obvious correlation between the SFR and the appearance of a gradient, although we cannot exclude that such a connection does exist but is obscured by high dust extinction. The average mean for those without gradients ($\overline{\rm SFR} = 3.6\pm1.1\,M_\odot\,{\rm yr}^{-1}$, where the uncertainty is the standard error) is slightly larger for those with a gradient ($\overline{\rm SFR} = 2.2\pm0.7\,M_\odot\,{\rm yr}^{-1}$), but given the large mean deviations, we cannot distinguish in a statistical sense between the two cases. This could be due to several factors. All of the gradients, except for those in NGC~1343 and NGC~4321, are partial gradients. That is, their gradient does not range continually from one contact point to the next. NGC~1343 and NGC~4321, which represent the``ideal'' scenario, do have high SFRs ($6.8\msunpyr$ for both cases) when compared to the rest of the sample. Computing the average SFR of the ring also is insensitive to the steepness of the gradient, the size of the ring, the size of the individual \hii\ regions, or the patchiness of the ring, which all add complexity to the interpretation of the data. 

From our analysis of age gradients we thus conclude that although age gradients along the ring are detected in more than half of the rings, all other rings show random variations of \hii\ region EW, and thus age, with azimuth. We cannot pinpoint any morphological parameters that might distinguish the host galaxies and bars of those rings with from those without gradients. Another strong conclusion from our analysis is that the kind of clear bi-polar age gradient seen before in NGC~4321 (Ryder, Knapen, \& Takamiya 2001; Allard et al. 2006) is rare - we only see a similar bi-polar gradient in one more galaxy, NGC~1343. It is not clear why these two galaxies show their gradients, whereas other galaxies with very similar bar-ring systems show gradients that are not bi-polar, or no gradients at all.

\subsection{Nuclear Ring and Stellar Bar Dynamics}

It is agreed that the presence of a galactic bar can have significant influence on angular momentum and re-distribution of gas, including inflow towards the central regions (e.g., Combes \& Gerin 1985; Athanassoula 1992, 2000; Knapen \ea\ 1995a; Heller \& Shlosman 1996; Maciejewski \ea\ 2002). Observed kinematics reveal that the shock-induced flows move gas radially inwards along the bar dust lanes and tend to collect at the intersection of the dust lanes with the ring. These contact points are thought to be approximately perpendicular (within the plane of the disk) to the PA of the bar major axis (Regan \& Teuben 2003) but have been seen to lead the bar by as little as 60$\deg$ in numerical models (Heller \& Shlosman 1996). The pile up of gas at these two points in the ring can be expected to spark SF. Therefore, one could predict to find the youngest stellar \hii\ regions in the ring near the contact points.

For the  barred spiral galaxies in our sample, we compare the location of the youngest \hii\ regions in the ring to the derived location of the contact points. We first identified the youngest \hii\ region in each hemisphere, where the hemispheres are separated by the bar PA. Taking into account the direction of flow around the ring (from our $B - I$ images), we calculate the PA offset between the \hii\ region and the contact points. Table~2 shows the PAs of both the contact point and the youngest \hii\ region in each hemisphere, along with the directional flow of the ring. We find that in two-thirds of the galaxies, one of the bar contact points is aligned with the youngest \hii\ region in the associated hemisphere within 20$^\circ$ of the contact point. Two galaxies, NGC~4314 and NGC~5248, have the youngest \hii\ region in both hemispheres aligned with the contact points. Our results for both of these rings are consistent with those of Benedict \ea\ (2002) and Maoz \ea\ (2001), respectively, who also find that the youngest stars were located at the contact points. Figure~10 shows a histogram of the distribution of the angular offsets between the bar major axis and the locations of the two youngest H{\sc ii} regions in each ring. A peak in frequency around the contact point locations is seen.

We next calculate the times from the bar major axis PA to the location of the contact points with those of the youngest \hii\ region in each hemisphere, assuming we azimuthally move in the direction of ring flow. We compute the period of the rings using measurements of the ring radius and the rotational velocity at the ring radius. We chose the major axis length of each ring as derived from {\sc ellipse} (see Section~3 and Table~1), and obtained the rotational velocity values from the literature and from our own DensePak IFU data taken in 2003 \& 2004 (L. M. Mazzuca \ea\ 2006, in prep.). Table~3 shows the major axis ring radius, the adopted rotational velocity along with its reference, and the period for each ring. Since we predict the contact points to be $\sim$90$^\circ$ from the bar major axis, we can compute their location with respect to the bar PA in each hemisphere, and thus calculate the amount of time it would take to travel to that location. We compare this travel time to that associated with the location of the youngest \hii\ region in each hemisphere.  Overall, the expected travel times to the contact points coincide with the times computed from the bar PA to the youngest \hii\ regions, which further strengthens the theory that the youngest \hii\ regions are located near the contact points. We find only one \hii\ region, in NGC~1530, that deviates from the general trend. In this case, the youngest region in one of the hemispheres is more aligned with the PA of the ring than that of the bar.

The analysis described above considers merely the rotational velocity of the gas in the ring region while deriving travel times. There is an additional contribution from the bar, which rotates with a certain bar pattern speed (assuming that the ring co-rotates with the bar). Estimating the relative contributions of these two rotation components, we consider parameters for a typical ring in our sample (see Table~1) of $r_{\rm ring}=0.5$\,kpc, $\Omega_{\rm p, bar}=40$\,km\,s$^{-1}$\,kpc$^{-1}$, and $v_{rot}=150$,km\,s$^{-1}$, which yields a ratio of the contributions of circular to bar pattern speeds into the velocities around the ring of $(40\times 0.5)/150$. Given that the rotational velocity is dominant, that the uncertainties in determining the bar pattern speed (and, in fact, the direction of bar rotation) are significant (see Elmegreen 1996 and Knapen 1999 for reviews), and that the uncertainties from other sources, as described above, in the determination of travel times around the ring are already rather large, we have not pursued to derive bar pattern speeds and incorporate them into our analysis.

\section{Conclusions}
From a larger \halpha\ imaging survey we identified 22 galaxies that contain a nuclear ring. To identify their sizes and shapes, we fitted each ring with an ellipse based on the peak isophotal intensity across the radial range of the ring. The rings vary in size from 0.2 kpc to 1.7 kpc in major axis radius. We compare the ring morphologies to those of their host galaxies. In all but one case the ellipticity and PA of the rings approximately match that of their respective disk, which corroborates the idea that the ring is in the same plane as the disk. The exception is NGC~7570, which shows a distinct and self-consistent difference in both the PA and eccentricity in the ring versus the disk. 

All but one of the rings had sufficient spatial resolution for us to identify individual \hii\ regions. For each \hii\ region, we computed the EW and converted differences between them into relative ages and age gradients using the Starburst99 evolutionary models. Although we adopted an instantaneous starburst model, we acknowledge that the rings may be more representative of a hybrid episodic scenario where star formation toggles between latent and active periods corresponding to the time it takes for gas build-up to reach critical density. Three rings, NGC~1343, NGC~1530, and NGC~4321, contain bipolar age gradients, where the ages increased from each of the contact points.  Many of the rings in our sample exhibit a well-defined age distribution pattern, but not throughout the entire ring. This can be a result of the combination of bar-induced dynamics and gravitational instabilities that are occurring on an intermittent basis. We calculated the SFR in each ring from the integrated luminosity contributed by all of the \hii\ regions forming the ring, and compared those rates to the appearance of a gradient within each ring (or lack thereof). The large uncertainties associated with variable extinction across the rings and the large dispersion of SFR values for both cases yields no clear correlation between the SFR and the emergence (or lack thereof) of an age-dependent azimuthal distribution. We do see a connection between the location of the bar contact points and the youngest \hii\ regions in two-thirds of the galaxies in our sample.

$acknowledgements  $
We thank Dr. Isaac Shlosman for inflow rate discussions and Dr. Torsten B{\" o}ker for input into earlier stages of the research. We also thank the referee for the thorough review of this paper. JHK acknowledges support from the Leverhulme Trust in the form of a Leverhulme Research Fellowship. The WHT is operated on the island
of La Palma by the Isaac Newton Group in the Spanish Observatorio del Roque de los Muchachos of the Instituto de Astrof\'\i sica de Canarias.  This research has made use of the NASA/IPAC Extragalactic Database (NED), which is operated by the Jet Propulsion Laboratory, California Institute of Technology, under contract with the National Aeronautics and Space Administration.

\clearpage
\begin{table} 
\centering
\begin{tabular}{lccccccccc}
\hline \hline
NGC & Morph & $\epsilon_{Ring}$ & $PA_{Ring}$ & $R_{Ring}$ & $R_{Ring}$ & $PA_{Disk}$ & $PA_{bar}$ & SFR\\
&  Type &  & ($\deg$) & ($\as$) & (kpc) & ($\deg$) & ($\deg$) & ($\msunpyr$)\\
(1) & (2) & (3) & (4) & (5) & (6) & (7) & (8) & (9) \\
\hline
~~278 & SAB(rs)b &  0.06 & 120 & 4.4 $\times$ 4.1 & 0.2 $\times$ 0.2& - &  no bar & 0.5\\
~~473 & SAB(r)0/a &  0.37 & 160 & 12.2 $\times$ 6.9 & 1.7 $\times$ 1.0& 153 & 164 & 2.2\\
~~613 & SB(rs)bc &  0.3 & 122 & 5.1 $\times$ 2.6 & 0.4 $\times$ 0.2& 120 & 120 & 2.2\\
1300 & (R)SB(s)bc &  0.25 & 135 & 4.1 $\times$ 3.1 & 0.3 $\times$ 0.2& 106& 102 &0.2\\
1343 & SAB(s)b &  0.25 & 60 & 8.8 $\times$ 6.6 & 1.2 $\times$ 0.9& 80 & 82 &6.8\\
1530 & SB(rs)b &  0.35 & 25 &6.8 $\times$ 4.9 & 1.2 $\times$ 0.8& 8 & 122 & 3.8\\
4303 & SAB(rs)bc &  0.14 & 88 & 3.3 $\times$ 2.8 & 0.2 $\times$ 0.2& - &10 &1.4\\
4314 & SB(rs)a &  0.10 & 135 & 6.6 $\times$ 5.9 & 0.3 $\times$ 0.3& - &135 &0.1\\
4321 & SAB(s)bc &  0.12 & 170 & 8.8 $\times$ 7.0 & 0.7 $\times$ 0.6& 30 & 153 & -\\
5248 & (R)SB(rs)bc &  0.3 & 115 & 6.6 $\times$ 4.6 & 0.7 $\times$ 0.5& 110 & 137 & 4.2\\
5728 & (R1)SAB(r)a & 0.4 & 125 & 5.3 $\times$ 3.2 & 1.1 $\times$ 0.6& - & 33 & 4.0\\
5905 & SB(r)b & 0.2 & 141 & 1.6 $\times$ 1.5 & 0.3 $\times$ 0.3& 135 &25 & 2.6\\
5945 & SB(rs)ab &  0.1 & 105 & 3.5 $\times$ 3.2 & 1.2 $\times$ 1.1& 105 &10 &4.4 \\
5953 & SAa &  0.1 & 192 & 6.1 $\times$ 5.5 & 1.0 $\times$ 0.9& 169 & no bar & 9.9\\
6503 & SA(s)cd &  0.65 & 121 & 38.5 $\times$ 13.4 & 1.1 $\times$ 0.4& 123 & no bar & 1.5\\
6951 & SAB(rs)bc &  0.2 & 146 & 4.6 $\times$ 3.7 & 0.5 $\times$ 0.4& 170 & 85 & 1.4\\
7217 & (R)SA(r)ab &  0.20 & 89 & 11 $\times$ 8.8 & 0.8 $\times$ 0.7& 95 & no bar & 0.6\\
IC1438 & SAB(r)a &  0.10 & 130 & 3.3 $\times$ 3.0	 & 0.5 $\times$ 0.5& - & 124 & 1.3\\
7469 & SAB (rs)a &  0.09 & 38 & 1.6 $\times$ 1.5 & 0.5 $\times$ 0.5 & 135 & 56 & -\\
7570 & SBa & 0.05 & 135 & 4.4 $\times$ 4.2 & 1.3 $\times$ 1.3& 30 & 30 & 1.4\\
7716 & SAB(r)b &  0.2 & 30 & 6.3 $\times$ 4.2 & 1.0 $\times$ 0.7& 35 & 34 &3.2\\
7742 & SA(r)b & 0.05 & 133 & 9.9 $\times$ 9.4 & 1.0 $\times$ 1.0& - & no bar& 4.3\\
\hline
\end{tabular}
\caption{Morphological characteristics of the galaxies in the observed sample. Galaxies are listed in order of increasing RA (Col.~1); morphological type as stated in the RC3 (Col.~2); ellipticity of the nuclear ring as derived by using the IRAF {\sc ellipse} task (Col.~3); PA of the ring major axis, calculated counterclockwise from North to East ({\sc ellipse}; Col.~4); ring radius in arcseconds (Col.~5) and kpc (Col.~6); PA of the galactic disk from RC3 (a dash represents a circular ring with no defined PA; Col.~7), except for NGC~1530 (Regan \ea\ 1996) which is based on kinematical data; PA of the bar major axis, except for NGC~1530 (Regan \ea\ 1996) and NGC~7469 (Davies, Tacconi, \& Genzel 2004) which are based on kinematical data (Col.~8); and the ring SFR (a dash represents lack of data necessary to calculate the SFR; Col.~9) is derived from the total ring luminosity using the Kennicutt (1998) \halpha-SFR relation.}
\end{table}

\clearpage

\begin{table}
\centering
\begin{tabular}{lccccccc}
\hline \hline
NGC & CP \#1 & \hii\ region \#1 & CP \#2 & \hii\ region \#2 & Rotation & Diff PA \#1 & Diff PA\#2\\
(1) & (2) & (3) & (4) & (5) & (6) & (7) & (8)\\
\hline
~473 & 254 & 312 & 74 & 38 & ccw & 58 & 36\\
~613 & 210 & 205 & 30 & 323 & ccw & 15 & 67\\
1300 & 12 & 345 & 192 & 269 & cw & 27 & 77\\
1343 & 172 & 140 & 352 & 309 & ccw & 32 & 43\\
1530 & 32 & 344 & 212 & 282 & cw & 48 & 70\\
4303 & 280 & 304 & 100 & 114 & cw & 24 & 14\\
4314 & 45 & 27 & 225 & 222 & cw & 18 & 3\\
5248 & 47 & 43 & 227 & 243 & ccw & 4 & 16\\
5728 & 303 & 305 & 123 & 147 & cw & 2 & 24 \\
5905 & 295 & 343 & 115 & 151 & cw & 48 & 36\\
5945 & 100 & 148 & 280 & 280 & ccw & 48 & 0\\
6951 & 355 & 339 & 175 & 100 & cw &  16 & 75\\
IC1438& 214 & 251 & 34 & 32 & ccw & 37 & 2\\
7570 & 300 & 353 & 120 & 131 & cw & 53 & 11\\
7716 & 124 & 91 & 304 & 301 & ccw & 33 & 3\\
\hline
\end{tabular}
\caption{Comparison of the location (i.e., PA) of the contact points (CPs) to that of the youngest \hii\ region in each hemisphere, as bisected from the bar. The PAs of each contact point, defined to occur 90$\deg$ offset from the bar PA, are given in Col.~2 and~4; the location of the youngest \hii\ region in each hemisphere is given in Col.~3 and~5 with the ring rotational direction cited in Col~6, as observed in the $B-I$ images; Col.~7 and~8 indicate the difference in location between the contact points and the respective \hii\ region. In most cases the separation is small, which is conistent with the notion that the youngest \hii\ region (in its respective hemisphere) should be near the location of the contact points.}
\end{table}

\clearpage

\begin{table} 
\centering
\begin{tabular}{lccccccc}
\hline \hline
NGC & Age  & $V_{\rm rot}$ & Ref & Period & CP time & \hii\ region \#1& \hii\ region \#2\\
& Distrib & ($\kms$) & & (Myr) & (Myr) & (Myr) & (Myr) \\
(1) & (2) & (3) & (4) & (5) & (6) & (7) & (8)  \\
\hline
~278 & az.gradient  & 63 & (a)&   35  & -     &  -    &  -       \\
~473 & az. gradient  & 127 & IFU & 103    & 72   & 89 & 62\\
~613 & flat & 115 & (b) & 23    & 13 & 13	& 9\\
1300 & no pattern & 135 & (c)& 17     & 0.6  &  16 & 4\\
1343 & az. gradient  & 131 & IFU & 39   & 18 & 15 & 14 \\
1530 & az. gradient  & 185 & IFU & 50   & 4 &  48 &14\\
4303 & no pattern & 97 & IFU & 9      & 6  &  8 &7\\
4314 & az. gradient & 161 & IFU & 30    & 4  & 2 &3\\
4321 & n/a &  170 & (d) & 32     & -	  &  -    & -      \\
5248 & az. gradient & 152 & IFU & 40   & 5    &5 &7 \\
5728 & az. gradient  & 180	& (e)& 47    & 45 &  40 &43\\
5905 & flat  & 150   & (f)& 15   & 11 &  14 &14\\
5945 & flat  & 150   & (g)& 62   & 17 &  25 &17\\
5953 & rad gradient  & 200 & IFU & 40  & -	  &  -    & -     \\
6503 & no pattern    & 90 & (h) & 94       & -	  &  -    & -      \\
6951 & az. gradient  & 160 & (c)& 24    & 24 &  23 &19 \\
7217 & az. gradient  & 248 & (i)& 25    & -	  &  -    &  -    \\
IC1438 & az. gradient  & 204 & (j) & 19  & 11 &  13  & 11\\
7570 & rad gradient  & 150 & (g) & 67& 66 & 65 &57\\
7716 & no pattern & 150 & (g) & 51       & 18 &  13 &17\\
7742 & no pattern  & 116 & IFU & 96  & -  &  -    &  -     \\
\hline
\end{tabular}
\caption{Age distribution and \hii\ region times. Col. 2 cites the age distribution, which fell into 4 categories: azimuthal gradient, radial gradient across the ring, flat, no pattern. For those rings with host bars, the period of the ring (Col. 5) is computed based on the rotational velocity at the ring radius (Col. 3) and the ring major axis radius (see Table~1, Col. 5);  literature references for the rotational velocites are in Col. 4: a) Knapen \ea\ (2004); b) Blackman (1981); c) Knapen, Laine, \& Rela\~no (1999); d) Knapen \ea\ (2000); e) Rubin (1980); f) Moiseev, Vald{\'e}s, \& Chavushyang (2004); g) no reference found- assumed a rotational velocity value of 150~$\kms$, which is the average of the measured values for the other rings (all rings in this sample have similar host galaxy morphologies and rotation curve characteristics associated with the ring radial location) ; h) Bottema (1989) for \hbeta line; i) Peterson \ea\ (1978); j) P. M. Treuthardt \& R. J. Buta, private communication; 'IFU' refers to our WIYN IFU data (L. M. Mazzuca et al. 2006). Travel times from the bar major axis PA to the contact points (Col~6) and to the youngest \hii\ region in each hemisphere (Col.~7 and~8, when a bar exists.}
\end{table}

\clearpage

\appendix
\section{Individual \hii\ Region Parameters for the Galaxies}
For each ring in the sample we present tables that contain each \hii\ region as identified by {\sc SExtractor} (col.~1), integrated flux (col.~2), luminosity (col.~3), EW (col~4) and log(EW) (col.~5). Column~6 indicates if the $I$-band flux needed to be modified in order to compensate for unrealistic values (i.e., negative flux values or marginally positive flux values -- see Section 4.2.1). In the cases where the {\it I}-band flux generated from {\sc SExtractor} was not modified, a value of '0' is recorded; otherwise a value of '1' is recorded. Since {\sc SExtractor} detects all \hii\ regions in the supplied image, extraneous \hii\ regions as well as the nucleus were included in the identification process. These superfluous detections are not included in the tables, but are instead noted in the captions for completeness sake. 

\clearpage

\begin{table} 
\begin{tabular}{lccccc}
\hline\hline
\hii\ region & Integrated Flux & Luminosity & EW & log(EW) & Flux \\ 
& (10$^{-13}$ \escm) & (10$^{39}$ \es) & (\AA) & (\AA) & Correction\\
(1) & (2) & (3) & (4) & (5) &(6) \\
\hline
1& 3.38& 5.63& 258& 2.41& 0\\ 
2& 0.98& 1.63& 229& 2.36& 1\\ 
3& 0.56& 0.93& 84& 1.92& 1\\ 
5& 1.74& 2.91& 485& 2.69& 1\\ 
7& 4.98& 8.30& 536& 2.73& 0\\ 
8& 0.52& 0.86& 81& 1.91& 1\\ 
9& 0.7& 1.17& 143& 2.15& 1\\ 
10& 6.20& 10.3& 3341& 3.52& 1\\ 
11& 0.51& 0.85& 4& 0.56& 0\\ 
12& 0.86& 1.44& 146& 2.16& 0\\ 
13& 0.66& 1.10& 128& 2.11& 1\\ 
14& 0.98& 1.63& 142& 2.15& 0\\ 
18& 3.36& 5.61& 365& 2.56& 0\\ 
19& 2.62& 4.36& 650& 2.81& 0\\ 
20& 1.12& 1.87& 284& 2.45& 1\\ 
21& 0.87& 1.45& 114& 2.06& 1\\ 
22& 0.54& 0.91& 53& 1.72& 1\\ 
23& 0.72& 1.20& 133& 2.12& 1\\ 
24& 0.90& 1.51& 222& 2.35& 1\\ 
25& 1.02& 1.71& 157& 2.20& 0\\ 
26& 3.54& 5.90& 1620& 3.21& 1\\ 
\hline
\end{tabular}
\caption{NGC~278 - 20 \hii\ regions detected associated with the ring. \hii\ regions \#~4, 6, and 11 comprise the nucleus; \hii\ regions \#~15, 16, and 17 are extraneous to the ring.}
\end{table}

\clearpage

\begin{table} 
\begin{tabular}{lccccc}
\hline\hline
\hii\ region & Integrated Flux & Luminosity & EW & log(EW) & Flux \\ 
& (10$^{-13}$ \escm) & (10$^{39}$ \es) & (\AA) & (\AA) & Correction\\
(1) & (2) & (3) & (4) & (5) &(6) \\
\hline
1& 0.33& 3.49& 216& 2.33& 1\\ 
4& 4.01& 42.6& 9343& 3.97& 1\\ 
5& 1.42& 15& 1643& 3.22& 0\\ 
6& 3.34& 35.5& 1409& 3.15& 0\\ 
7& 1.88& 20& 2660& 3.42& 0\\ 
8& 0.62& 6.57& 608& 2.78& 1\\ 
9& 0.35& 3.73& 222& 2.35& 1\\ 
10& 0.37& 3.93& 262& 2.42& 1\\ 
11& 1.03& 10.9& 942& 2.97& 0\\ 
12& 1.15& 12.3& 1812& 3.26& 1\\ 
13& 0.42& 4.44& 389& 2.59& 1\\ 
14& 0.22& 2.37& 122& 2.09& 1\\ 
15& 1.32& 14.1& 647& 2.81& 0\\ 
16& 0.11& 1.20& 41& 1.62& 1\\ 
17& 1.13& 12& 1722& 3.24& 1\\ 
18& 0.60& 6.37& 647& 2.81& 1\\ 
19& 0.32& 3.37& 200& 2.30& 1\\ 
20& 1.60& 17& 2241& 3.35& 0\\ 
21& 2.08& 22.1& 2622& 3.42& 0\\ 
22& 0.83& 8.85& 1138& 3.06& 1\\ 
25& 1.55& 16.5& 1261& 3.10& 0\\ 
26& 0.17& 1.83& 70& 1.85& 1\\ 
28& 0.32& 3.36& 192& 2.28& 1\\ 
29& 0.10& 1.05& 25& 1.41& 1\\ 
30& 0.22& 2.32& 120& 2.08& 1\\ 
32& 0.19& 2.02& 78& 1.89& 1\\ 
33& 0.26& 2.79& 107& 2.03& 1\\ 
\hline
\end{tabular}
\caption{NGC~473 - 27 \hii\ regions detected associated with the ring. \hii\ regions \#~23 and 24 comprise the nucleus; \hii\ regions \#~2, 3, 27, and 31 are extraneous to the ring.}

\end{table}
\clearpage

\begin{table} 
\begin{tabular}{lccccc}
\hline\hline
\hii\ region & Integrated Flux & Luminosity & EW & log(EW) & Flux \\ 
& (10$^{-13}$ \escm) & (10$^{39}$ \es) & (\AA) & (\AA) & Correction\\
(1) & (2) & (3) & (4) & (5) &(6) \\
\hline
1& 26.3& 96.4& 179& 2.25& 0\\ 
3& 13.3& 48.9& 88& 1.95& 0\\ 
4& 3.87& 14.2& 189& 2.28& 0\\ 
5& 2.26& 8.28& 374& 2.57& 0\\ 
6& 8.49& 31.1& 26& 1.42& 0\\ 
7& 1.00& 3.66& 130& 2.12& 0\\ 
8& 2.00& 7.33& 221& 2.34& 0\\ 
9& 0.84& 3.08& 77& 1.88& 0\\ 
10& 0.86& 3.14& 368& 2.57& 0\\ 
12& 11.4& 41.8& 148& 2.17& 0\\ 
13& 5.89& 21.6& 87& 1.94& 0\\ 
\hline
\end{tabular}
\caption{NGC~613 - 12 \hii\ regions detected. \hii\ regions \#~2 and 11 are extraneous to the ring.}
\end{table}
\vspace{0in}


\begin{table} 
\begin{tabular}{lccccc}
\hline\hline
\hii\ region & Integrated Flux & Luminosity & EW & log(EW) & Flux \\ 
& (10$^{-13}$ \escm) & (10$^{39}$ \es) & (\AA) & (\AA) & Correction\\
(1) & (2) & (3) & (4) & (5) &(6) \\
\hline
1& 1.34& 5.69& 107& 2.03& 0\\ 
2& 1.34& 5.67& 56& 1.75& 0\\ 
3& 0.21& 0.88& 47& 1.67& 1\\ 
4& 0.05& 0.23& 7& 0.85& 1\\ 
5& 0.40& 1.71& 118& 2.07& 1\\ 
6& 0.16& 0.67& 33& 1.52& 1\\ 
7& 0.15& 0.63& 25& 1.39& 1\\ 
8& 0.18& 0.75& 37& 1.57& 1\\ 
10& 0.70& 2.95& 240& 2.38& 1\\ 
11& 0.51& 2.17& 155& 2.19& 1\\ 
12& 0.45& 1.91& 158& 2.20& 1\\ 
13& 0.39& 1.64& 56& 1.74& 0\\ 
\hline
\end{tabular}
\caption{NGC~1300 - 12 \hii\ regions detected associated with the ring. \hii\ region \#~9 is the nucleus.}
\end{table}
\vspace{0in}

\clearpage

\begin{table} 
\begin{tabular}{lccccc}
\hline\hline
\hii\ region & Integrated Flux & Luminosity & EW & log(EW) & Flux \\ 
& (10$^{-13}$ \escm) & (10$^{39}$ \es) & (\AA) & (\AA) & Correction\\
(1) & (2) & (3) & (4) & (5) &(6) \\
\hline
1& 8.24& 85.8& 5364& 3.73& 1\\ 
2& 10.2& 106& 744& 2.87& 0\\ 
3& 4.73& 49.2& 330& 2.52& 0\\ 
4& 2.14& 22.3& 247& 2.39& 0\\ 
5& 3.88& 40.4& 650& 2.81& 0\\ 
6& 1.36& 14.2& 396& 2.60& 1\\ 
7& 1.44& 15& 400& 2.60& 1\\ 
8& 18.9& 196& 5845& 3.77& 0\\ 
9& 0.36& 3.79& 33& 1.52& 1\\ 
10& 0.65& 6.73& 80& 1.91& 1\\ 
11& 1.37& 14.3& 401& 2.60& 1\\ 
12& 3.95& 41.1& 737& 2.87& 0\\ 
16& 2.55& 26.5& 968& 2.99& 1\\ 
17& 4.61& 48& 373& 2.57& 0\\ 
18& 3.22& 33.6& 1340& 3.13& 1\\ 
19& 4.12& 42.9& 1339& 3.13& 0\\ 
20& 0.72& 7.53& 117& 2.07& 1\\ 
21& 4.84& 50.4& 2513& 3.40& 1\\ 
22& 3.77& 39.3& 479& 2.68& 0\\ 
23& 1.52& 15.8& 305& 2.48& 1\\ 

\hline
\end{tabular}
\caption{NGC~1343 - 20 \hii\ regions detected associated with the ring. \hii\ regions \#~13 and 14 comprise the nucleus; \hii\ region \#~15 is extraneous to the ring.}
\end{table}
\vspace{.3in}
\clearpage

\begin{table} 
\begin{tabular}{lccccc}
\hline\hline
\hii\ region & Integrated Flux & Luminosity & EW & log(EW) & Flux \\ 
& (10$^{-13}$ \escm) & (10$^{39}$ \es) & (\AA) & (\AA) & Correction\\
(1) & (2) & (3) & (4) & (5) &(6) \\
\hline
1& 6.79& 109& 1895& 3.28& 0\\ 
2& 13.4& 215& 630& 2.80& 0\\ 
3& 1.46& 23.4& 243& 2.38& 0\\ 
4& 1.05& 16.8& 232& 2.37& 0\\ 
7& 0.52& 8.32& 66& 1.82& 0\\ 
11& 0.52& 8.31& 854& 2.93& 1\\ 
14& 1.84& 29.5& 274& 2.44& 0\\ 
15& 0.98& 15.8& 1951& 3.29& 1\\ 
17& 1.55& 24.8& 3123& 3.49& 1\\ 

\hline
\end{tabular}
\caption{NGC~1530 - 9 \hii\ regions detected associated with the ring. \hii\ regions \#~8, 9, 10, 12, 13, and 16 comprise the nucleus; \hii\ regions \#~5 and 6 are extraneous to the ring.}
\end{table}
\vspace{.3in}

\begin{table} 
\begin{tabular}{lccccc}
\hline\hline
\hii\ region & Integrated Flux & Luminosity & EW & log(EW) & Flux \\ 
& (10$^{-13}$ \escm) & (10$^{39}$ \es) & (\AA) & (\AA) & Correction\\
(1) & (2) & (3) & (4) & (5) &(6) \\
\hline
2& 25.7& 71.1& 99& 2.00& 0\\ 
3& 17.8& 49.2& 112& 2.05& 0\\ 
4& 1.33& 3.67& 367& 2.57& 0\\ 
5& 6.82& 18.9& 624& 2.80& 0\\ 
6& 1.35& 3.74& 74& 1.87& 0\\ 
7& 9.61& 26.6& 1676& 3.22& 0\\ 
8& 1.14& 3.14& 106& 2.03& 0\\ 

\hline
\end{tabular}
\caption{NGC~4303 - 7 \hii\ regions detected associated with the ring. \hii\ region \#~1 comprises the nucleus.}
\end{table}
\vspace{.3in}
\clearpage
 

\begin{table} 
\begin{tabular}{lccccc}
\hline\hline
\hii\ region & Integrated Flux & Luminosity & EW & log(EW) & Flux \\ 
& (10$^{-13}$ \escm) & (10$^{39}$ \es) & (\AA) & (\AA) & Correction\\
(1) & (2) & (3) & (4) & (5) &(6) \\
\hline
1& 3.32& 3.74& 529& 2.72& 0\\ 
2& 3.78& 4.25& 2607& 3.42& 0\\ 
3& 0.39& 0.44& 111& 2.05& 0\\ 
4& 0.23& 0.26& 148& 2.17& 0\\ 
5& 2.13& 2.40& 375& 2.57& 0\\ 
7& 2.82& 3.18& 1801& 3.26& 0\\ 
8& 0.12& 0.14& 344& 2.54& 0\\ 
9& 0.54& 0.61& 211& 2.32& 0\\ 
10& 0.42& 0.48& 1088& 3.04& 0\\ 
11& 0.46& 0.52& 1288& 3.11& 0\\ 

\hline
\end{tabular}
\caption{NGC~4313 - 10 \hii\ regions detected associated with the ring. \hii\ region \#~6 comprises the nucleus.}
\end{table}
\vspace{.3in}


\begin{table} 
\begin{tabular}{lccccc}
\hline\hline
\hii\ region & Integrated Flux & Luminosity & EW & log(EW) & Flux \\ 
& (10$^{-13}$ \escm) & (10$^{39}$ \es) & (\AA) & (\AA) & Correction\\
(1) & (2) & (3) & (4) & (5) &(6) \\
\hline
2& 27.8& 171& 1106& 3.04& 0\\ 
3& 4.08& 25.2& 259& 2.41& 0\\ 
4& 1.23& 7.58& 206& 2.31& 0\\ 
5& 0.92& 5.68& 1277& 3.11& 1\\ 
6& 1.82& 11.2& 6790& 3.83& 1\\ 
7& 6.18& 38.1& 121& 2.08& 0\\ 
8& 2.88& 17.8& 222& 2.35& 0\\ 
9& 1.06& 6.53& 49& 1.69& 0\\ 
10& 3.76& 23.2& 6378& 3.80& 0\\ 
11& 1.87& 11.5& 2275& 3.36& 0\\ 
12& 4.93& 30.4& 485& 2.69& 0\\ 
13& 2.03& 12.5& 539& 2.73& 0\\ 
14& 7.44& 45.9& 191& 2.28& 0\\ 
15& 14& 86.1& 965& 2.98& 0\\ 
16& 0.57& 3.52& 78& 1.89& 0\\ 
17& 1.86& 11.5& 153& 2.19& 0\\ 
18& 1.51& 9.33& 3125& 3.49& 1\\ 
19& 0.88& 5.40& 1145& 3.06& 1\\ 
20& 1.82& 11.2& 179& 2.25& 0\\ 

\hline
\end{tabular}
\caption{NGC~5248 - 19 \hii\ regions detected associated with the ring. \hii\ region \#~1 comprises the nucleus.}
\end{table}
\vspace{.3in}
\clearpage


\begin{table} 
\begin{tabular}{lccccc}
\hline\hline
\hii\ region & Integrated Flux & Luminosity & EW & log(EW) & Flux \\ 
& (10$^{-13}$ \escm) & (10$^{39}$ \es) & (\AA) & (\AA) & Correction\\
(1) & (2) & (3) & (4) & (5) &(6) \\
\hline
3& 2.28& 48.6& 1164& 3.07& 0\\ 
4& 0.27& 5.76& 163& 2.21& 0\\ 
5& 2.13& 45.4& 7961& 3.90& 1\\ 
6& 2.19& 46.7& 248& 2.40& 0\\ 
7& 3.87& 82.5& 556& 2.74& 0\\ 
8& 0.21& 4.48& 170& 2.23& 1\\ 
9& 0.70& 15& 1516& 3.18& 1\\ 
10& 0.16& 3.53& 109& 2.04& 1\\ 
11& 0.30& 6.32& 479& 2.68& 1\\ 
12& 0.17& 3.58& 111& 2.04& 1\\ 
13& 0.22& 4.78& 226& 2.35& 1\\ 
14& 0.62& 13.3& 1218& 3.09& 1\\ 
15& 1.11& 23.6& 3084& 3.49& 1\\ 
17& 4.09& 87.2& 211& 2.32& 0\\ 
18& 3.52& 75& 1426& 3.15& 0\\ 
19& 1.53& 32.6& 1310& 3.12& 0\\ 
21& 0.24& 5.10& 182& 2.26& 1\\ 
22& 0.39& 8.38& 455& 2.66& 0\\ 

\hline
\end{tabular}
\caption{NGC~5728 - 18 \hii\ regions detected associated with the ring. \hii\ regions \#~1 and 2 comprise the nucleus; \hii\ regions \#~16 and 20 are extraneous to the ring.}
\end{table}
\vspace{.3in}
\clearpage

\begin{table} 
\begin{tabular}{lccccc}
\hline\hline
\hii\ region & Integrated Flux & Luminosity & EW & log(EW) & Flux \\ 
& (10$^{-13}$ \escm) & (10$^{39}$ \es) & (\AA) & (\AA) & Correction\\
(1) & (2) & (3) & (4) & (5) &(6) \\
\hline
1& 2.39& 58.5& 77& 1.88& 0\\ 
2& 2.63& 64.3& 189& 2.28& 0\\ 
3& 0.40& 9.80& 97& 1.99& 0\\ 
4& 0.67& 16.3& 2870& 3.46& 1\\ 
5& 0.41& 10& 177& 2.25& 0\\ 
6& 0.38& 9.37& 154& 2.19& 0\\ 
7& 0.25& 6.20& 199& 2.30& 0\\ 
8& 3.27& 79.9& 87& 1.94& 0\\ 
9& 1.75& 42.7& 132& 2.12& 0\\ 
\hline
\end{tabular}
\caption{NGC~5905 - 9 \hii\ regions detected associated with the ring. No nuclear or extraneous \hii\ regions detected.}
\end{table}
\vspace{.3in}


\begin{table} 
\begin{tabular}{lccccc}
\hline\hline
\hii\ region & Integrated Flux & Luminosity & EW & log(EW) & Flux \\ 
& (10$^{-13}$ \escm) & (10$^{39}$ \es) & (\AA) & (\AA) & Correction\\
(1) & (2) & (3) & (4) & (5) &(6) \\
\hline
2& 3.84& 249& 15& 1.17& 0\\ 
3& 3.82& 248& 16& 1.21& 0\\ 
4& 0.93& 60.2& 29& 1.46& 0\\ 
\hline
\end{tabular}
\caption{NGC~5945 - 3 \hii\ regions detected associated with the ring. \hii\ region \#~1 comprises the nucleus.}
\end{table}
\vspace{.3in}
\clearpage


\begin{table} 
\begin{tabular}{lccccc}
\hline\hline
\hii\ region & Integrated Flux & Luminosity & EW & log(EW) & Flux \\ 
& (10$^{-13}$ \escm) & (10$^{39}$ \es) & (\AA) & (\AA) & Correction\\
(1) & (2) & (3) & (4) & (5) &(6) \\
\hline
2& 11.6& 152& 152& 2.18& 0\\ 
3& 9.84& 128& 390& 2.59& 0\\ 
4& 24.9& 324& 541& 2.73& 0\\ 
5& 2.24& 29.2& 652& 2.81& 1\\ 
6& 1.22& 15.9& 247& 2.39& 1\\ 
7& 6.38& 83.2& 614& 2.79& 0\\ 
8& 2.56& 33.3& 591& 2.77& 0\\ 
9& 5.41& 70.5& 2162& 3.33& 1\\ 
10& 4.35& 56.7& 784& 2.89& 0\\ 
11& 2.19& 28.5& 94& 1.97& 0\\ 
12& 3.22& 42& 1013& 3.01& 1\\ 
13& 0.43& 5.57& 30& 1.48& 1\\ 
14& 0.74& 9.67& 88& 1.94& 1\\ 
15& 2.01& 26.1& 4& 0.61& 0\\ 
16& 0.45& 5.82& 4& 0.55& 0\\ 
17& 2.57& 33.5& 130& 2.12& 0\\ 
18& 1.86& 24.3& 50& 1.70& 0\\ 
19& 3.25& 42.4& 154& 2.19& 0\\ 
20& 3.63& 47.4& 1374& 3.14& 1\\ 
21& 7.02& 91.5& 866& 2.94& 0\\ 
22& 0.33& 4.30& 23& 1.37& 1\\ 
23& 0.46& 5.99& 48& 1.68& 1\\ 

\hline
\end{tabular}
\caption{NGC~5953 - 22 \hii\ regions detected associated with the ring. \hii\ region \#~1 comprises the nucleus.}
\end{table}
\vspace{.3in}
\clearpage


\begin{table} 
\begin{tabular}{lccccc}
\hline\hline
\hii\ region & Integrated Flux & Luminosity & EW & log(EW) & Flux \\ 
& (10$^{-13}$ \escm) & (10$^{39}$ \es) & (\AA) & (\AA) & Correction\\
(1) & (2) & (3) & (4) & (5) &(6) \\
\hline
3& 38.6& 17.2& 3686& 3.57& 0\\ 
4& 7.12& 3.17& 804& 2.91& 1\\ 
5& 1.08& 0.48& 45& 1.65& 1\\ 
6& 2.28& 1.01& 151& 2.18& 1\\ 
7& 1.55& 0.69& 52& 1.71& 1\\ 
8& 1.90& 0.84& 67& 1.83& 1\\ 
9& 3.43& 1.53& 224& 2.35& 1\\ 
10& 2.08& 0.93& 98& 1.99& 1\\ 
11& 4.07& 1.81& 307& 2.49& 1\\ 
12& 0.94& 0.42& 26& 1.42& 1\\ 
13& 10.7& 4.76& 1715& 3.23& 1\\ 
14& 2.60& 1.16& 214& 2.33& 1\\ 
15& 2.64& 1.18& 120& 2.08& 1\\ 
16& 1.13& 0.50& 40& 1.60& 1\\ 
17& 0.55& 0.25& 12& 1.08& 1\\ 
18& 0.59& 0.26& 13& 1.11& 1\\ 
19& 1.00& 0.45& 31& 1.49& 1\\ 
20& 10.5& 4.66& 796& 2.90& 0\\ 
21& 14.8& 6.59& 1123& 3.05& 0\\ 
22& 12.3& 5.46& 2072& 3.32& 1\\ 
23& 4.66& 2.08& 304& 2.48& 1\\ 
24& 0.87& 0.39& 22& 1.34& 1\\ 
25& 2.64& 1.18& 190& 2.28& 1\\ 
26& 1.32& 0.59& 70& 1.85& 1\\ 
27& 9.70& 4.32& 1271& 3.10& 1\\ 
28& 1.22& 0.54& 37& 1.57& 1\\ 
29& 0.95& 0.42& 21& 1.32& 1\\ 
30& 5.72& 2.55& 431& 2.63& 1\\ 
31& 2.87& 1.28& 148& 2.17& 1\\ 
32& 2.32& 1.03& 92& 1.96& 1\\ 
33& 6.78& 3.02& 808& 2.91& 1\\ 
34& 2.18& 0.97& 132& 2.12& 1\\ 
35& 3.92& 1.75& 323& 2.51& 1\\ 
36& 11.8& 5.26& 1774& 3.25& 1\\ 
37& 33& 14.7& 6628& 3.82& 1\\ 

\hline
\end{tabular}
\caption{NGC~6503 - 84 \hii\ regions detected associated with the ring. \hii\ regions \#~39, 44, and 45 comprise the nucleus; \hii\ regions \#~1, 2, and 86 are extraneous to the ring.}
\end{table}
\clearpage

\setcounter{table}{16}
\begin{table} 
\begin{tabular}{lccccc}
\hline\hline
\hii\ region & Integrated Flux & Luminosity & EW & log(EW) & Flux \\ 
& (10$^{-13}$ \escm) & (10$^{39}$ \es) & (\AA) & (\AA) & Correction\\
(1) & (2) & (3) & (4) & (5) &(6) \\
\hline
38& 5.41& 2.41& 584& 2.77& 1\\ 
40& 5.55& 2.47& 570& 2.76& 1\\ 
41& 9.64& 4.29& 1331& 3.12& 1\\ 
42& 7.17& 3.19& 661& 2.82& 1\\ 
43& 2.88& 1.28& 240& 2.38& 1\\ 
46& 9.87& 4.39& 1363& 3.13& 1\\ 
47& 1.36& 0.61& 51& 1.71& 1\\ 
48& 1.30& 0.58& 49& 1.69& 1\\ 
49& 3.55& 1.58& 299& 2.48& 1\\ 
50& 1.85& 0.82& 66& 1.82& 0\\ 
51& 1.19& 0.53& 40& 1.60& 1\\ 
52& 3.29& 1.47& 278& 2.44& 1\\ 
53& 3.95& 1.76& 337& 2.53& 1\\ 
54& 2.51& 1.12& 191& 2.28& 1\\ 
55& 2.97& 1.32& 129& 2.11& 1\\ 
56& 2.92& 1.30& 172& 2.24& 1\\ 
57& 10.3& 4.57& 1594& 3.20& 1\\ 
58& 6.08& 2.71& 657& 2.82& 1\\ 
59& 7.01& 3.12& 682& 2.83& 1\\ 
60& 1.16& 0.51& 44& 1.64& 1\\ 
61& 2.40& 1.07& 150& 2.18& 1\\ 
62& 2.72& 1.21& 16& 1.19& 0\\ 
63& 20.8& 9.27& 2839& 3.45& 0\\ 
64& 0.75& 0.34& 19& 1.28& 1\\ 
65& 1.17& 0.52& 47& 1.67& 1\\ 
66& 8.01& 3.57& 1106& 3.04& 1\\ 
67& 0.62& 0.28& 14& 1.13& 1\\ 
68& 1.49& 0.66& 65& 1.81& 1\\ 
69& 37.2& 16.6& 7731& 3.89& 1\\ 
70& 1.54& 0.69& 51& 1.71& 1\\ 
71& 2.34& 1.04& 110& 2.04& 1\\ 
72& 1.57& 0.70& 71& 1.85& 1\\ 
73& 1.49& 0.66& 49& 1.69& 1\\ 
74& 1.69& 0.75& 60& 1.78& 1\\ 
75& 2.83& 1.26& 174& 2.24& 1\\ 
76& 1.65& 0.74& 75& 1.87& 1\\ 
77& 0.36& 0.16& 10& 1.00& 1\\ 

\hline
\end{tabular}
\caption{NGC~6503 - continued.}
\end{table}
\vspace{.3in}
\clearpage

\setcounter{table}{16}
\begin{table} 
\begin{tabular}{lccccc}
\hline\hline
\hii\ region & Integrated Flux & Luminosity & EW & log(EW) & Flux \\ 
& (10$^{-13}$ \escm) & (10$^{39}$ \es) & (\AA) & (\AA) & Correction\\
(1) & (2) & (3) & (4) & (5) &(6) \\
\hline
78& 1.80& 0.80& 78& 1.89& 1\\ 
79& 2.94& 1.31& 236& 2.37& 1\\ 
80& 3.84& 1.71& 312& 2.49& 1\\ 
81& 3.86& 1.72& 219& 2.34& 0\\ 
82& 1.16& 0.52& 62& 1.79& 1\\ 
83& 1.38& 0.61& 81& 1.91& 1\\ 
84& 4.53& 2.02& 483& 2.68& 1\\ 
85& 1.27& 0.57& 45& 1.65& 1\\ 
87& 2.03& 0.90& 117& 2.07& 1\\ 
88& 4.13& 1.84& 356& 2.55& 1\\ 
89& 2.45& 1.09& 106& 2.03& 1\\ 
90& 1.14& 0.51& 35& 1.55& 1\\ 

\hline
\end{tabular}
\caption{NGC~6503 - continued.}
\end{table}
\vspace{.3in}
\clearpage

\begin{table} 
\begin{tabular}{lccccc}
\hline\hline
\hii\ region & Integrated Flux & Luminosity & EW & log(EW) & Flux \\ 
& (10$^{-13}$ \escm) & (10$^{39}$ \es) & (\AA) & (\AA) & Correction\\
(1) & (2) & (3) & (4) & (5) &(6) \\
\hline
2& 7.63& 53& 58& 1.77& 0\\ 
3& 0.60& 4.15& 120& 2.08& 1\\ 
4& 2.30& 16& 810& 2.91& 1\\ 
5& 0.87& 6.08& 95& 1.98& 0\\ 
6& 8.92& 62& 3034& 3.48& 0\\ 
7& 1.36& 9.45& 154& 2.19& 0\\ 
8& 2.30& 16& 103& 2.01& 0\\ 
9& 1.38& 9.60& 37& 1.56& 0\\ 
\hline
\end{tabular}
\caption{NGC~6951 - 8 \hii\ regions detected associated with the ring. \hii\ region \#~1 comprises the nucleus.}
\end{table}
\vspace{.3in}
\clearpage

\begin{table} 
\begin{tabular}{lccccc}
\hline\hline
\hii\ region & Integrated Flux & Luminosity & EW & log(EW) & Flux \\ 
& (10$^{-13}$ \escm) & (10$^{39}$ \es) & (\AA) & (\AA) & Correction\\
(1) & (2) & (3) & (4) & (5) &(6) \\
\hline
2& 1.70& 5.20& 88& 1.94& 0\\ 
3& 1.06& 3.25& 542& 2.73& 1\\ 
6& 0.43& 1.32& 118& 2.07& 1\\ 
8& 0.65& 1.99& 224& 2.35& 1\\ 
9& 0.73& 2.23& 306& 2.49& 1\\ 
10& 0.50& 1.54& 153& 2.18& 1\\ 
11& 1.84& 5.64& 987& 2.99& 1\\ 
12& 0.79& 2.43& 263& 2.42& 1\\ 
13& 0.61& 1.86& 216& 2.33& 1\\ 
14& 10.3& 31.5& 56& 1.74& 0\\ 
15& 0.40& 1.23& 105& 2.02& 1\\ 
16& 0.26& 0.80& 42& 1.62& 1\\ 
17& 0.76& 2.32& 262& 2.42& 0\\ 
18& 0.21& 0.64& 33& 1.52& 1\\ 
19& 0.52& 1.59& 143& 2.16& 1\\ 
20& 0.64& 1.95& 137& 2.14& 0\\ 
21& 0.62& 1.90& 189& 2.28& 1\\ 
22& 0.23& 0.71& 48& 1.68& 1\\ 
23& 1.06& 3.25& 478& 2.68& 1\\ 
24& 0.77& 2.37& 246& 2.39& 1\\ 
25& 0.31& 0.94& 63& 1.80& 1\\ 
26& 1.19& 3.64& 477& 2.68& 1\\ 
\hline
\end{tabular}
\caption{NGC~7217 - 22 \hii\ regions detected associated with the ring. \hii\ regions \#~1 and 4 comprise the nucleus; \hii\ regions \#~5 and 7 are extraneous to the ring.}
\end{table}
\vspace{.3in}
\clearpage


\begin{table} 
\begin{tabular}{lccccc}
\hline\hline
\hii\ region & Integrated Flux & Luminosity & EW & log(EW) & Flux \\ 
& (10$^{-13}$ \escm) & (10$^{39}$ \es) & (\AA) & (\AA) & Correction\\
(1) & (2) & (3) & (4) & (5) &(6) \\
\hline
1& 0.49& 6.77& 56& 1.75& 0\\ 
2& 1.11& 15.2& 1054& 3.02& 1\\ 
3& 0.24& 3.27& 71& 1.85& 1\\ 
4& 0.65& 8.96& 65& 1.81& 0\\ 
5& 0.91& 12.4& 685& 2.84& 1\\ 
6& 0.31& 4.30& 86& 1.94& 1\\ 
9& 2.21& 30.2& 582& 2.77& 0\\ 
10& 3.38& 46.2& 234& 2.37& 0\\ 
11& 1.30& 17.8& 43& 1.63& 0\\ 
12& 0.95& 13& 244& 2.39& 0\\ 
13& 0.39& 5.31& 49& 1.69& 0\\ 
\hline
\end{tabular}
\caption{IC~1438 - 11 \hii\ regions detected associated with the ring. \hii\ regions \#~7 and 8 comprise the nucleus.}
\end{table}
\vspace{.3in}
\clearpage

\begin{table} 
\begin{tabular}{lccccc}
\hline\hline
\hii\ region & Integrated Flux & Luminosity & EW & log(EW) & Flux \\ 
& (10$^{-13}$ \escm) & (10$^{39}$ \es) & (\AA) & (\AA) & Correction\\
(1) & (2) & (3) & (4) & (5) &(6) \\
\hline
1& 0.49& 12& 446& 2.65& 0\\ 
2& 0.25& 6.11& 234& 2.37& 1\\ 
3& 0.35& 8.46& 410& 2.61& 1\\ 
4& 0.41& 10& 439& 2.64& 1\\ 
5& 0.16& 3.85& 93& 1.97& 1\\ 
6& 0.77& 18.9& 915& 2.96& 1\\ 
7& 0.44& 10.7& 343& 2.54& 1\\ 
8& 0.48& 11.7& 374& 2.57& 1\\ 
9& 1.43& 35& 3371& 3.53& 1\\ 
14& 0.17& 4.26& 73& 1.86& 1\\ 
15& 0.31& 7.49& 203& 2.31& 1\\ 
16& 0.12& 3.04& 74& 1.87& 1\\ 
17& 0.21& 5.18& 199& 2.30& 1\\ 
18& 0.38& 9.42& 302& 2.48& 1\\ 
19& 0.30& 7.45& 146& 2.17& 0\\ 
20& 0.25& 6.08& 217& 2.34& 0\\ 
21& 0.14& 3.38& 92& 1.96& 1\\ 
22& 0.80& 19.6& 1009& 3.00& 1\\ 
\hline
\end{tabular}
\caption{NGC~7570 - 18 \hii\ regions detected associated with the ring. \hii\ regions \#~10, 11, 12, and 13 comprise the nucleus.}
\end{table}
\vspace{.3in}
\clearpage

\begin{table} 
\begin{tabular}{lccccc}
\hline\hline
\hii\ region & Integrated Flux & Luminosity & EW & log(EW) & Flux \\ 
& (10$^{-13}$ \escm) & (10$^{39}$ \es) & (\AA) & (\AA) & Correction\\
(1) & (2) & (3) & (4) & (5) &(6) \\
\hline
1& 3.14& 42.6& 1516& 3.18& 0\\ 
2& 1.53& 20.7& 684& 2.84& 0\\ 
3& 1.38& 18.8& 436& 2.64& 0\\ 
4& 0.75& 10.2& 1259& 3.10& 0\\ 
5& 0.30& 4.13& 320& 2.51& 0\\ 
6& 0.52& 7.02& 1643& 3.22& 1\\ 
7& 0.18& 2.48& 304& 2.48& 1\\ 
8& 0.31& 4.24& 526& 2.72& 0\\ 
10& 0.5& 6.80& 1478& 3.17& 1\\ 
11& 0.84& 11.4& 2556& 3.41& 0\\ 
12& 0.40& 5.40& 643& 2.81& 0\\ 
13& 0.31& 4.19& 694& 2.84& 1\\ 
14& 1.20& 16.4& 4475& 3.65& 1\\ 
15& 0.10& 1.41& 100& 2.00& 1\\ 
17& 2.98& 40.6& 1840& 3.26& 0\\ 
18& 0.95& 13& 2904& 3.46& 0\\ 
19& 3.42& 46.5& 20204& 4.31& 1\\ 
20& 0.63& 8.57& 316& 2.50& 0\\ 
21& 0.83& 11.3& 102& 2.01& 0\\ 
22& 0.64& 8.71& 2039& 3.31& 1\\ 
23& 2.06& 28& 1135& 3.06& 0\\ 
24& 0.20& 2.76& 413& 2.62& 1\\ 
25& 0.56& 7.64& 1146& 3.06& 0\\ 
26& 0.50& 6.83& 1523& 3.18& 1\\ 
29& 0.29& 3.93& 519& 2.71& 1\\ 
30& 0.69& 9.32& 1100& 3.04& 0\\ 
32& 0.27& 3.64& 575& 2.76& 1\\ 
33& 0.45& 6.08& 1323& 3.12& 1\\ 
34& 3.28& 44.6& 23884& 4.38& 1\\ 
35& 0.59& 8.08& 1756& 3.24& 1\\ 
\hline
\end{tabular}
\caption{NGC~7716 - 30 \hii\ regions detected associated with the ring. \hii\ region \#~16 comprises the nucleus; \hii\ regions \#~9, 27, 28, and 31 are extraneous to the ring.}
\end{table}
\clearpage


\begin{table} 
\begin{tabular}{lccccc}
\hline\hline
\hii\ region & Integrated Flux & Luminosity & EW & log(EW) & Flux \\ 
& (10$^{-13}$ \escm) & (10$^{39}$ \es) & (\AA) & (\AA) & Correction\\
(1) & (2) & (3) & (4) & (5) &(6) \\
\hline
1& 1.70& 10& 254& 2.41& 0\\ 
2& 6.64& 39.2& 352& 2.55& 0\\ 
3& 6.59& 38.9& 484& 2.68& 0\\ 
4& 3.61& 21.3& 524& 2.72& 0\\ 
5& 1.08& 6.37& 284& 2.45& 1\\ 
6& 1.84& 10.9& 496& 2.70& 0\\ 
7& 0.88& 5.19& 236& 2.37& 1\\ 
8& 12.4& 73.2& 2057& 3.31& 0\\ 
9& 0.61& 3.59& 129& 2.11& 1\\ 
10& 1.84& 10.9& 696& 2.84& 1\\ 
11& 0.81& 4.80& 151& 2.18& 1\\ 
12& 2.28& 13.5& 1003& 3.00& 1\\ 
13& 0.99& 5.82& 347& 2.54& 1\\ 
14& 5.07& 29.9& 674& 2.83& 0\\ 
15& 1.45& 8.54& 507& 2.70& 1\\ 
16& 0.70& 4.15& 191& 2.28& 1\\ 
17& 0.39& 2.30& 55& 1.74& 1\\ 
19& 3.09& 18.2& 464& 2.67& 0\\ 
20& 1.68& 9.90& 575& 2.76& 0\\ 
21& 5.59& 32.9& 3721& 3.57& 1\\ 
22& 1.05& 6.21& 317& 2.50& 1\\ 
23& 2.55& 15& 254& 2.40& 0\\ 
24& 2.51& 14.8& 274& 2.44& 0\\ 
25& 1.63& 9.59& 626& 2.80& 1\\ 
26& 0.27& 1.59& 27& 1.43& 1\\ 
27& 2.83& 16.7& 677& 2.83& 0\\ 
28& 1.14& 6.75& 342& 2.53& 1\\ 
29& 0.94& 5.54& 244& 2.39& 1\\ 
30& 1.17& 6.88& 234& 2.37& 1\\ 
31& 2.76& 16.3& 1300& 3.11& 1\\ 
32& 0.42& 2.47& 84& 1.92& 1\\ 
33& 0.58& 3.45& 122& 2.09& 1\\ 
34& 2.00& 11.8& 811& 2.91& 1\\ 
35& 0.54& 3.19& 116& 2.06& 1\\ 
\hline
\end{tabular}
\caption{NGC~7742 - 38 \hii\ regions detected associated with the ring. \hii\ region \#~18 comprises the nucleus.}
\end{table}
\clearpage

\setcounter{table}{22}
\begin{table} 
\begin{tabular}{lccccc}
\hline\hline
\hii\ region & Integrated Flux & Luminosity & EW & log(EW) & Flux \\ 
& (10$^{-13}$ \escm) & (10$^{39}$ \es) & (\AA) & (\AA) & Correction\\
(1) & (2) & (3) & (4) & (5) &(6) \\
\hline
36& 5.52& 32.6& 3043& 3.48& 1\\ 
37& 1.22& 7.19& 390& 2.59& 1\\ 
38& 0.38& 2.25& 68& 1.83& 1\\ 
39& 4.75& 28& 2633& 3.42& 1\\ 
\hline
\end{tabular}
\caption{NGC~7742 continued.}
\end{table}

\clearpage

\begin{figure*}
\setcounter{figure}{0}
\epsscale{1}
\plotone{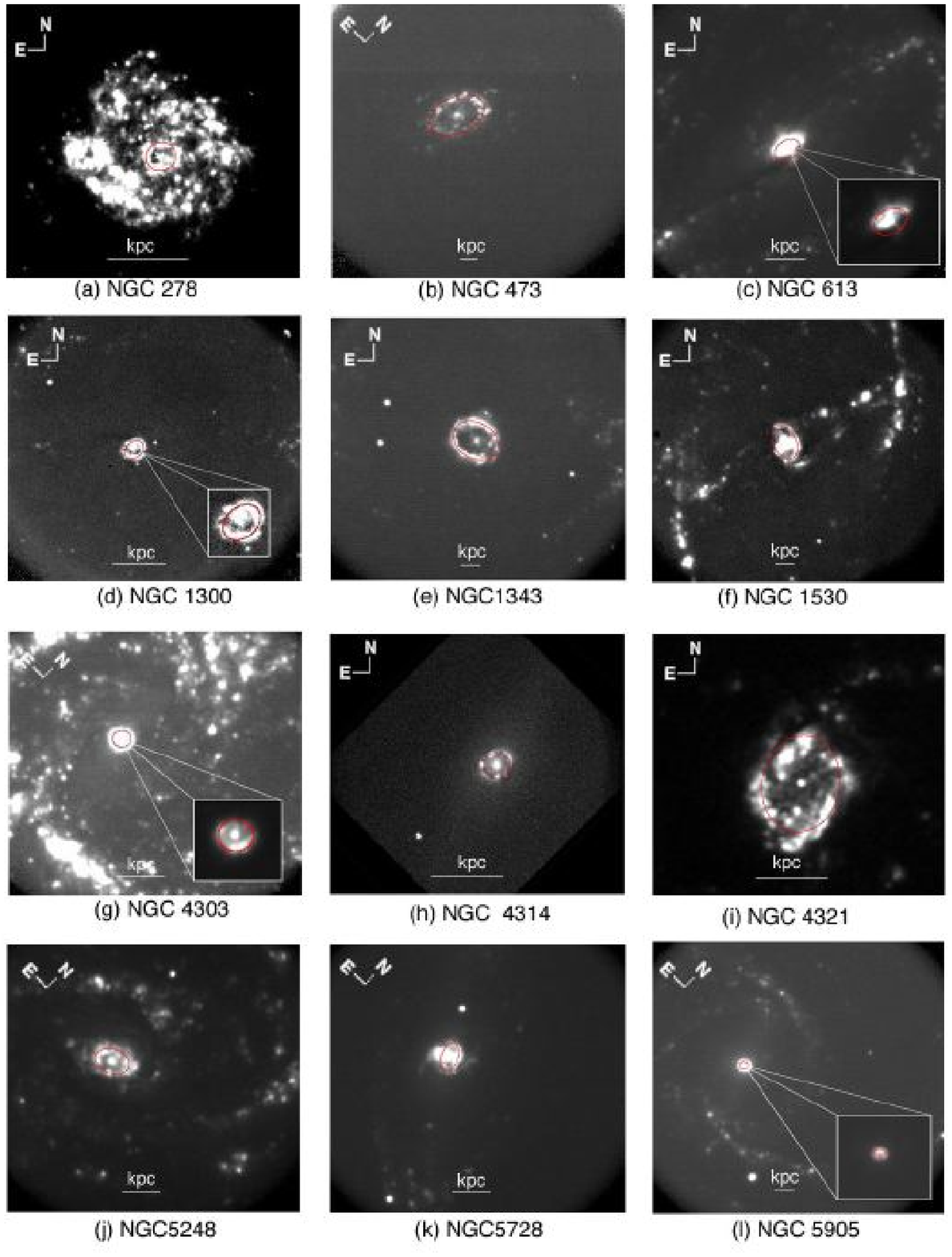}
\caption{panels a. - l.: Ellipse fits for the nuclear rings in the sample galaxies 
produced by the IRAF task {\sc ellipse}, overlaid on the \halpha\ images. The
major axis lengths were optimized based on the maximum intensity along a given
contour as computed using {\sc ellipse} (see Section~3).}
\end{figure*}
\clearpage 

\begin{figure*}
\setcounter{figure}{0}
\epsscale{1}
\plotone{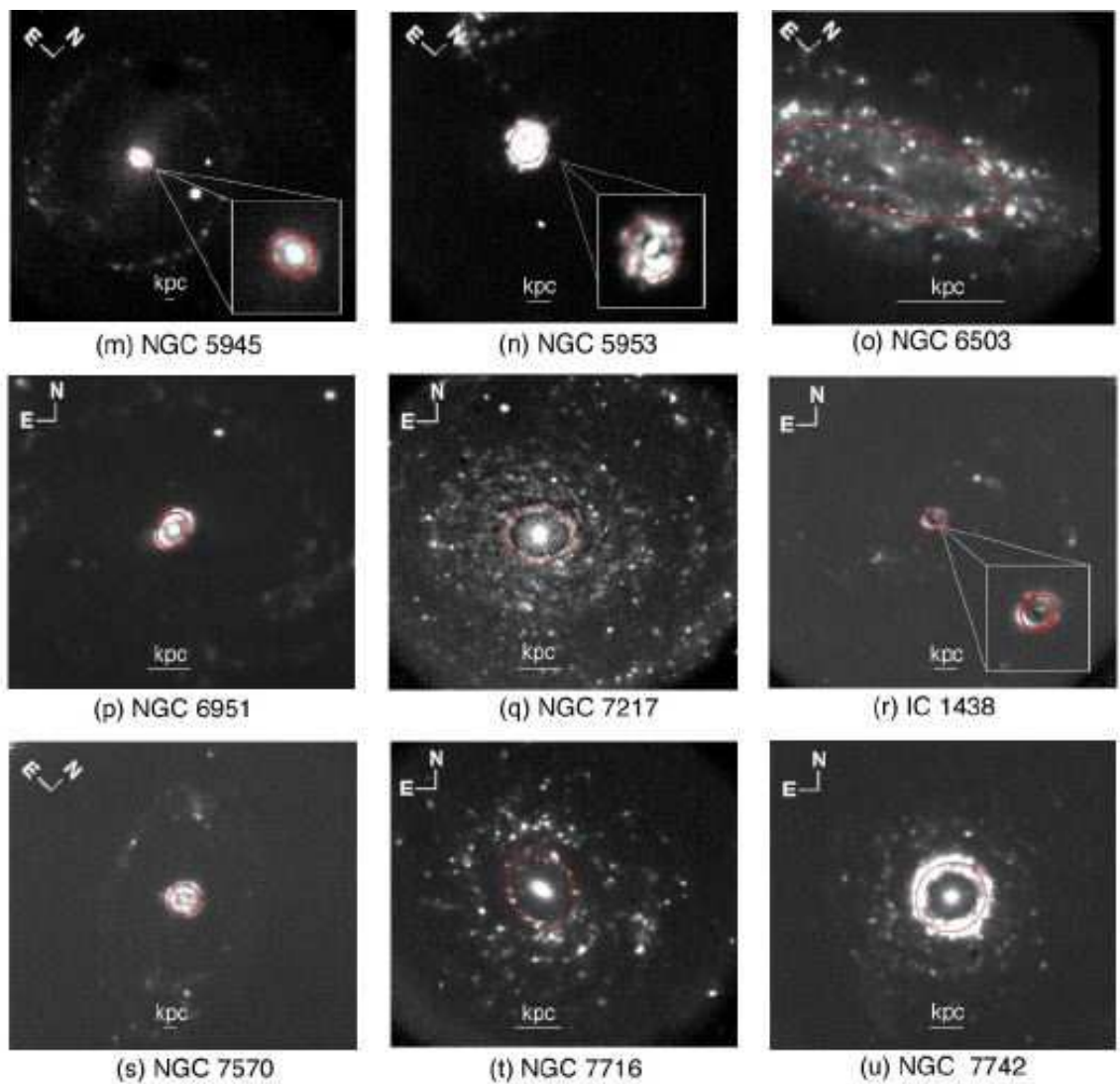}
\caption{continued: panels m. - v.}
\end{figure*}
\clearpage

\begin{figure*}
\setcounter{figure}{1}
\epsscale{1.0}
\plotone{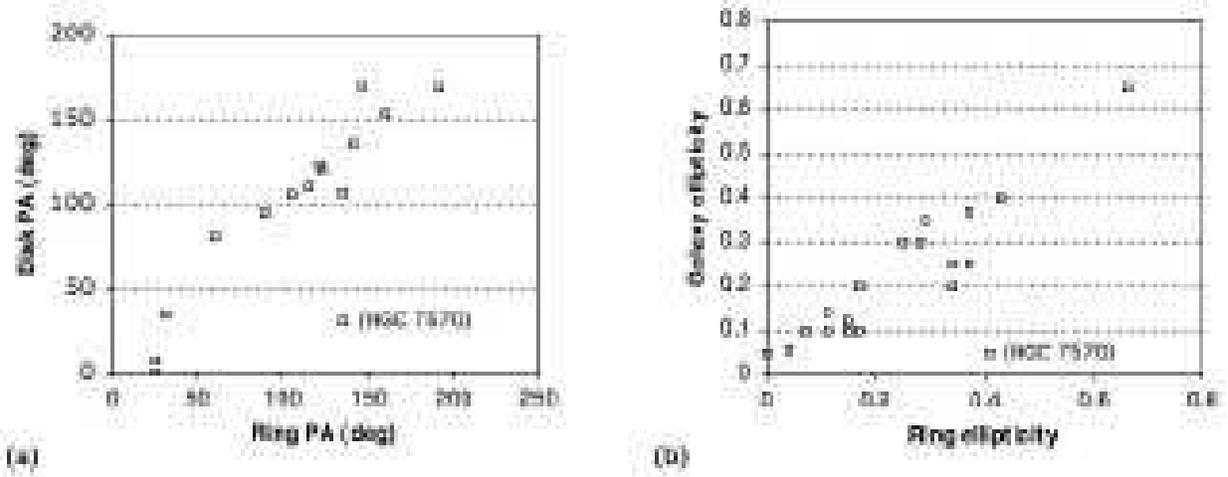}
\caption{(a) PA of the disk versus the ring major axis for those rings whose ellipticity is greater than or equal to 0.1. (b) Ellipticity comparison between the ring and the galactic disk. Both plots show a linear relationship between the ring and host galaxy PA and ellipticity, and indicate that the rings are in the same plane as the disk and are nearly circular.}
\end{figure*}

\begin{figure*}
\setcounter{figure}{2}
\epsscale{1}
\plotone{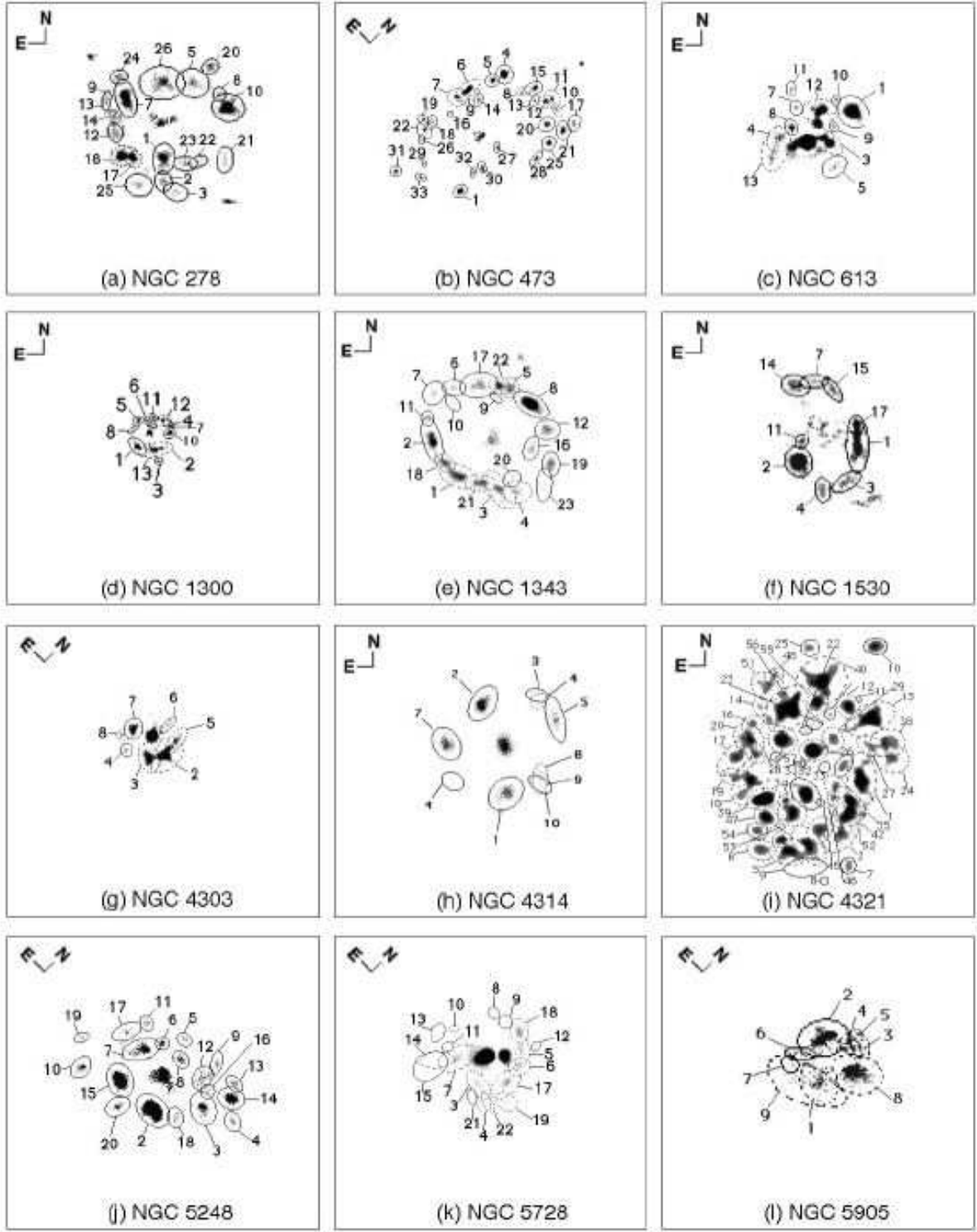}
\caption{panels a. - l.: Results from {\sc SExtractor}. Each \hii\ region identified is fitted using a Kron ellipse, which calculates the total magnitude of the extended \hii\ region. In the cases where the \halpha\ boundaries are merged from one or more sources, the program deblends the regions and indicates these fittings with a dashed ellipse. {\sc SExtractor} could not identify any \hii\ regions for NGC~7469 due to poor spatial resolution.}
\end{figure*}
\clearpage 

\begin{figure*}
\setcounter{figure}{2}
\epsscale{1}
\plotone{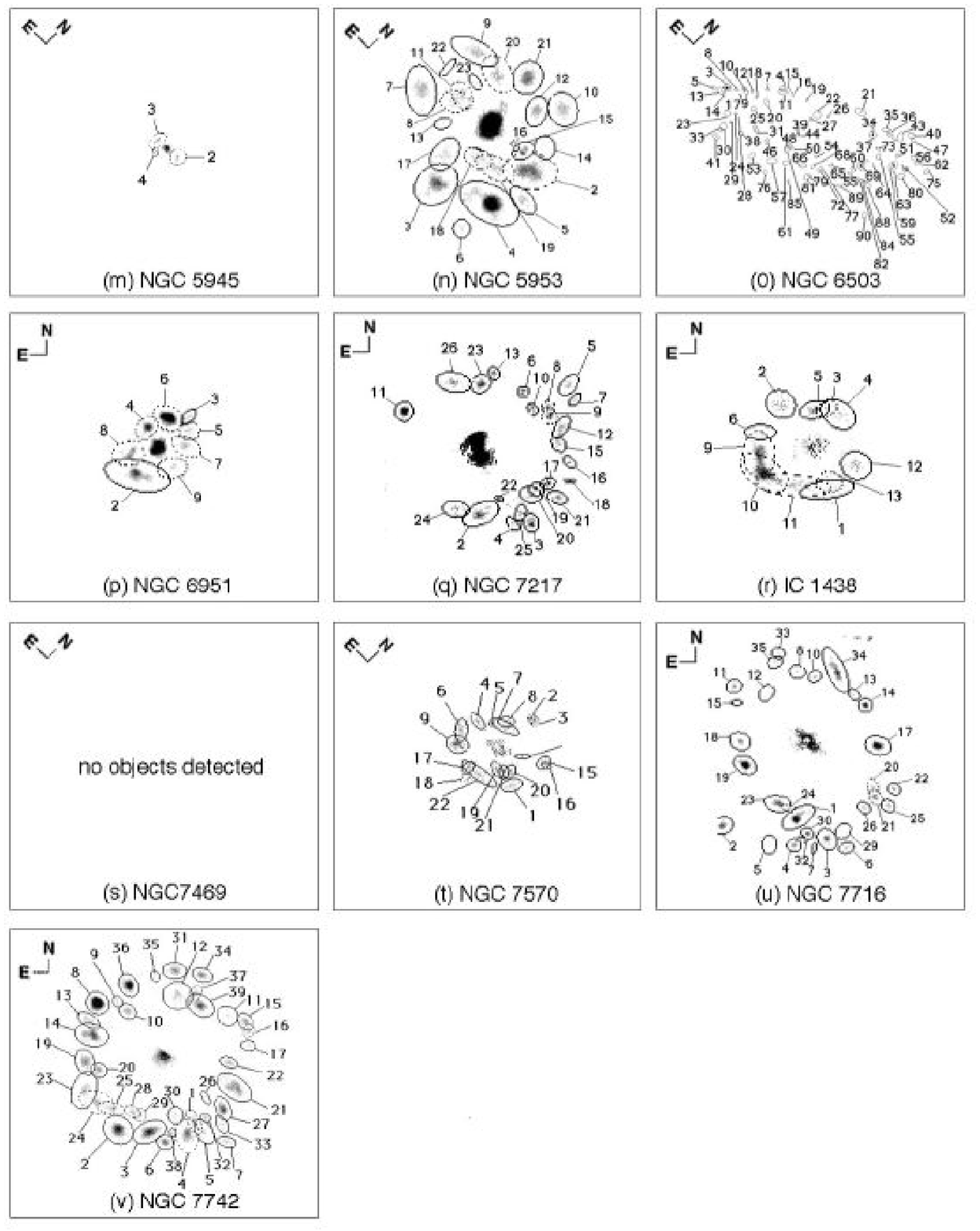}
\end{figure*}
\clearpage 

\begin{figure*}
\setcounter{figure}{3}
\epsscale{0.65}
\plotone{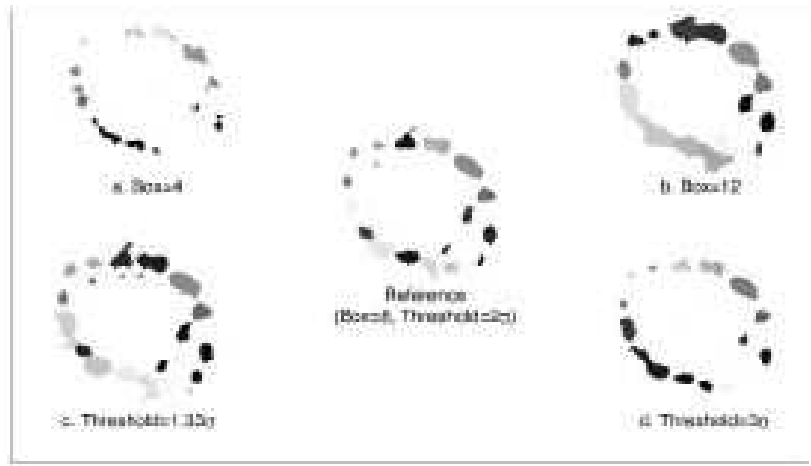}
\caption{Change in ring morphology for NGC 1343 based on variation of the background box size and detection threshold value. Top row shows the resolved \hii\ regions based on fixing the threshold at 2$\sigma$ and varying the  box size - a) box size = 4x4 pixels with 31 \hii\ region detections; b) box size = 12x12 pixels with 20 detections.  The center image is the reference image: box size = 8x8 pixels with 25 detections. Bottom row shows the images based on fixing the box size at 8x8 pixels and varying the threshold value - c) threshold = 1.33$\sigma$ with 30 detections; d) threshold = 3$\sigma$ with 19 detections. The optimal solution was derived iteratively and based on comparison of the general size, shape, and quantity of the regions that could be easily detected by eye in the Halpha image. The grayscale is $not$ representative of the ages of the regions, but rather the order in which {\sc SExtractor} identified the regions.}
\end{figure*}
\vspace{.2in}

\begin{figure*}
\setcounter{figure}{4}
\epsscale{0.85}
\plotone{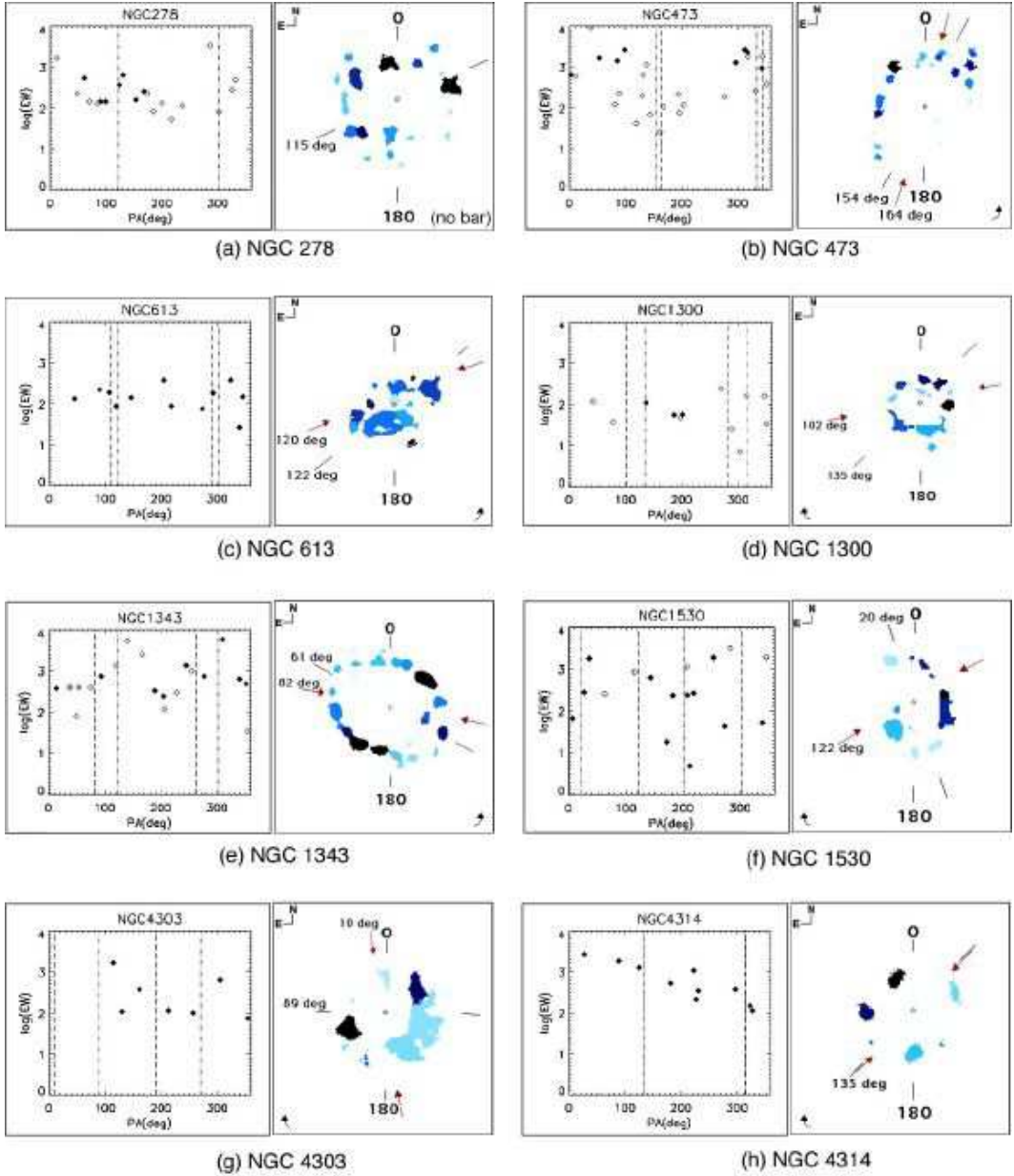}
\caption{Age distribution -- a. through h. The left panels plot the PA of each \hii\ region vs. their EW. Solid diamonds represent those EWs whose {\it I}-band fluxes were not affected by the background 3$\sigma$ noise; open diamonds indicate EW where {\it I}-band fluxes needed to be increased to compensate for the dominant background noise; open diamonds with crosses show those \hii\ regions that would not have needed their flux modified if using a 2$\sigma$ constraint. The right panels picturize the rings with their identified \hii\ regions. Age decreases as the shade of color changes from light to dark.  Arrows are drawn to indicate the PA of the bar major axis, as well as the rotational direction of flow in the ring (when a bar exists). The ring major axis is indicated by two dashed lines. All rings have been rotated to a typical North-East configuration. See Section~6.1 for details. NGC~4321 is based on \hbeta\ analysis from Allard \ea\ (2006).}
\end{figure*}

\begin{figure*}
\setcounter{figure}{4}
\epsscale{1}
\plotone{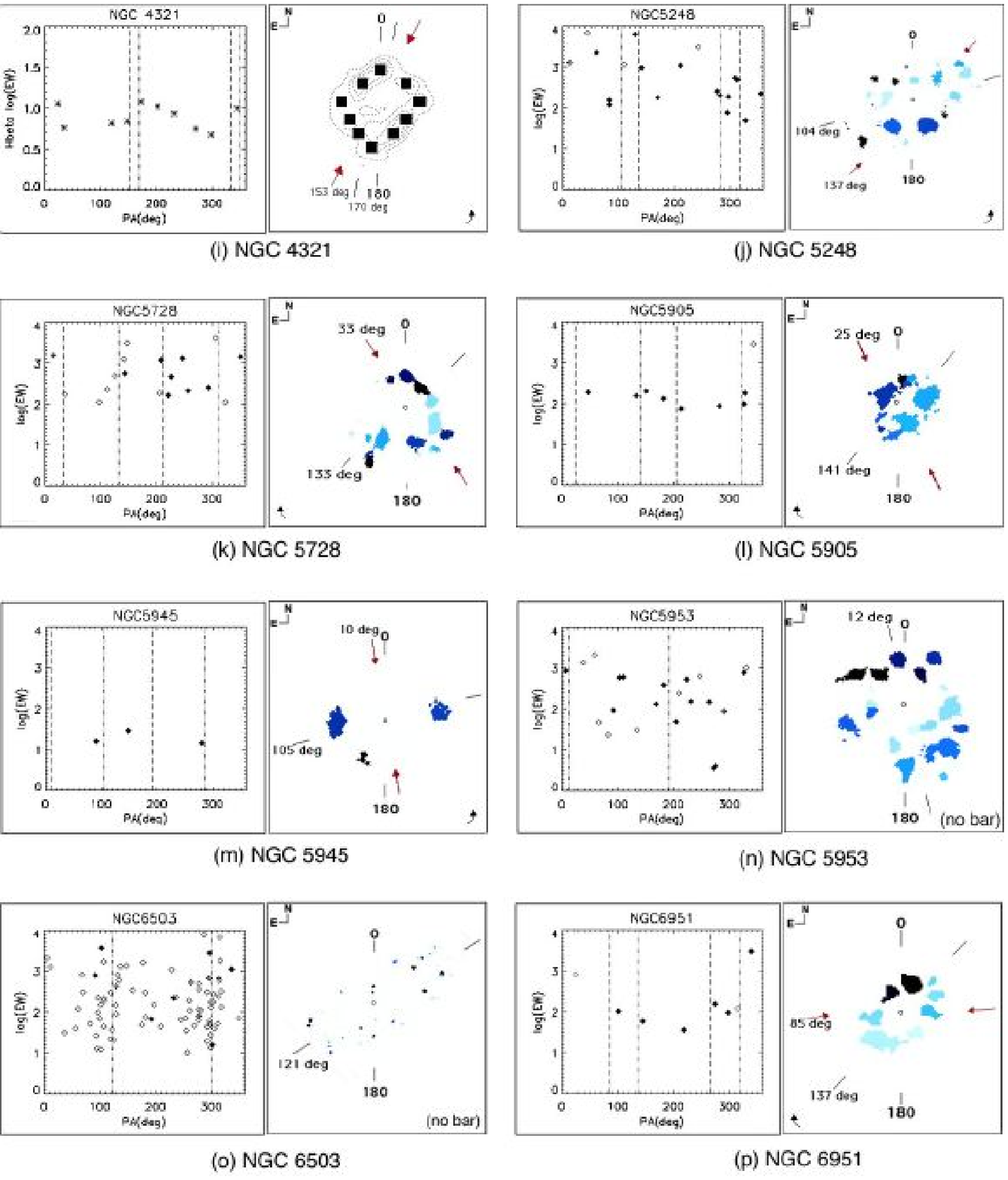}
\end{figure*}

\begin{figure*}
\setcounter{figure}{4}
\epsscale{1}
\plotone{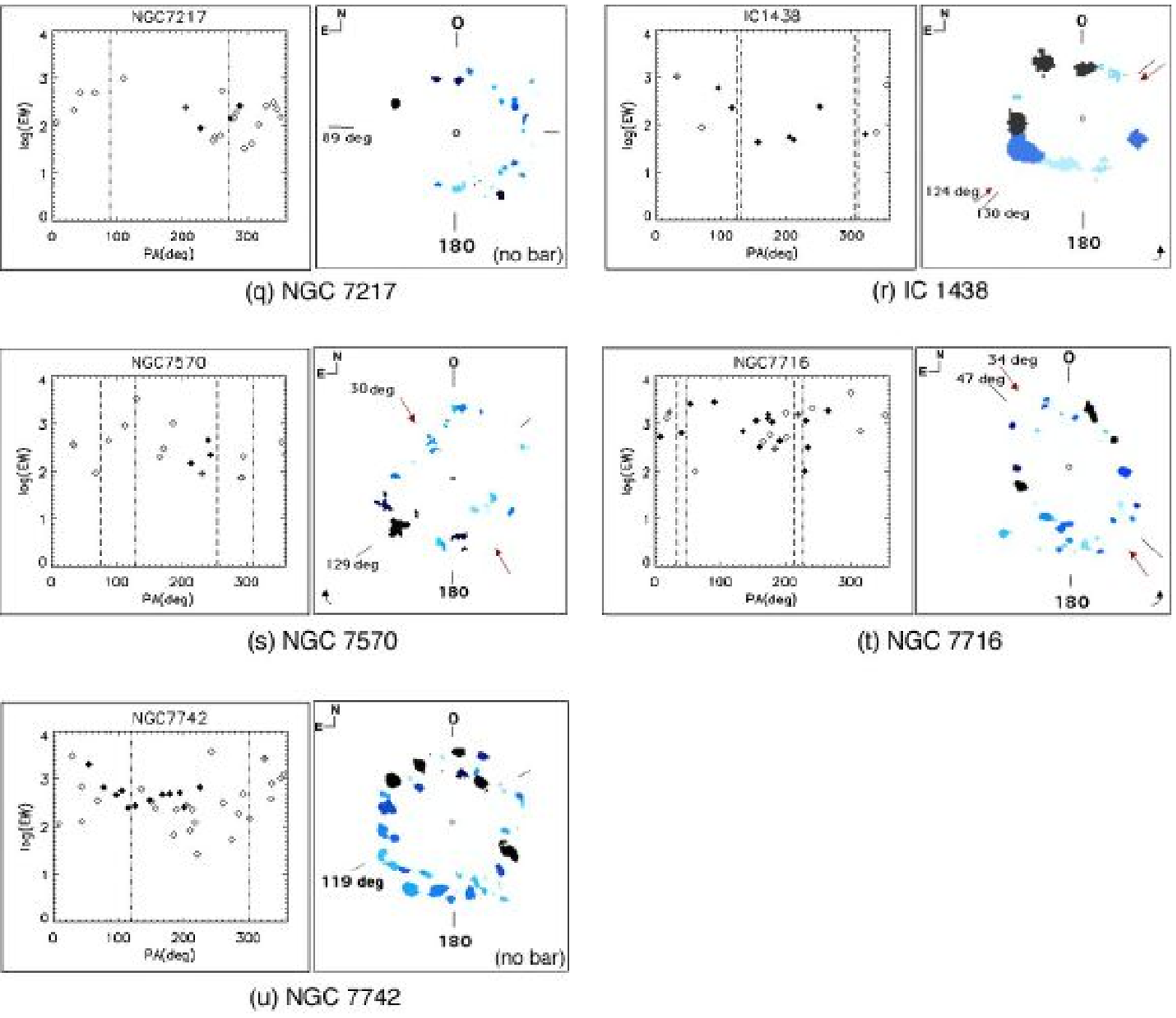}
\end{figure*}

\begin{figure*}
\setcounter{figure}{5}
\epsscale{0.5}
\plotone{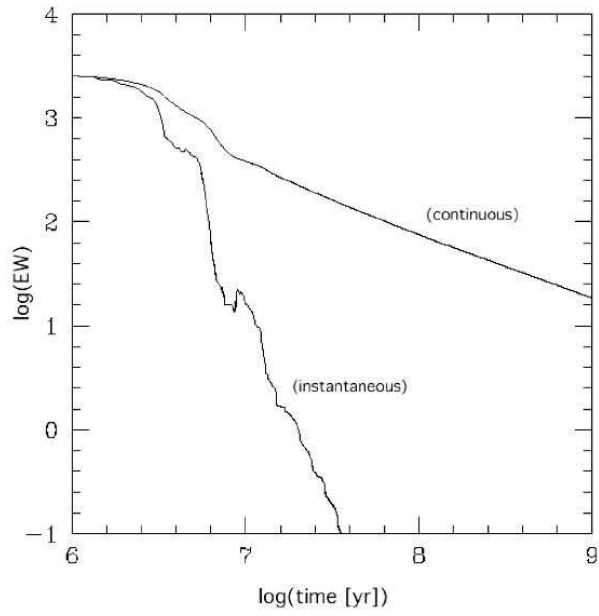}
\caption{\halpha\ EW vs. time following Leitherer et al. (1999; Starburst99 curves) for an instantaneous and a continuous burst of star formation with a Salpeter IMF between $M_{low} = 1\msun$ and $M_{up} = 100\msun$. A solar metallicity of $Z$ = 0.02 is used.}
\end{figure*}
\vspace{.2in}

\begin{figure*}
\setcounter{figure}{6}
\epsscale{1}
\plotone{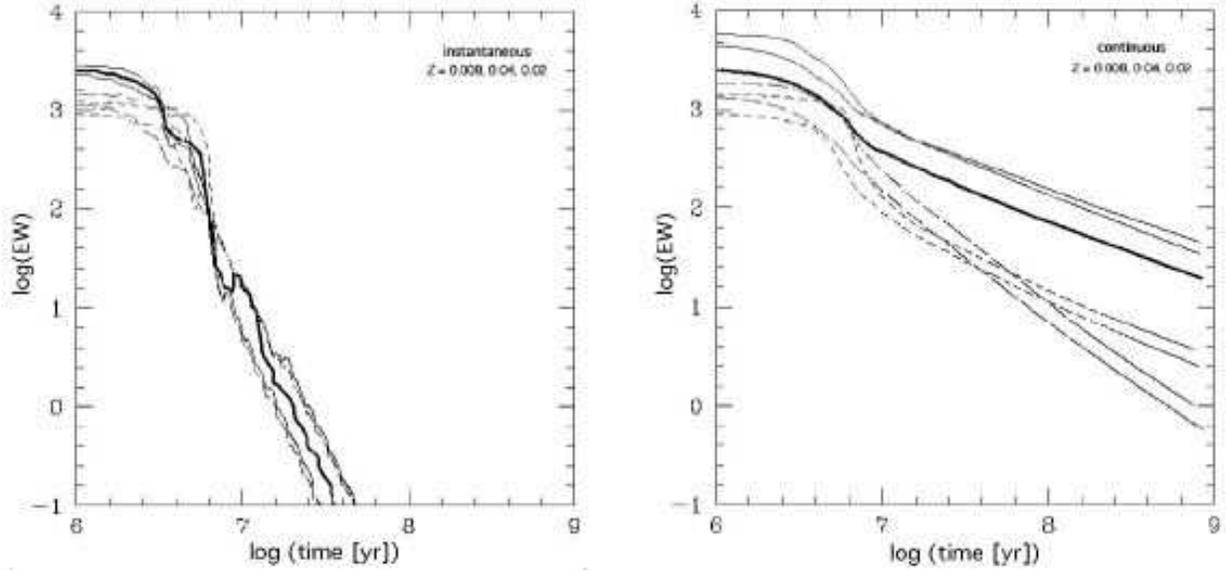}
\caption{\halpha\ EW vs. time following Leitherer et al. (1999; Starburst99 curves) for an instantaneous burst of star formation (left panel) and continuous burst (right panel) for varying IMF slopes, upper limit masses, and metalicities. The thick solid line represents the model prediction we adopted for IMF between $M_{low} = 1\msun$ and $M_{up} = 100\msun$, IMF slope= 2.35, and $Z$ = 0.02. Long dashed lines represent a slope of 3.3 and $M_{up} = 100\msun$ for $Z$=0.04 and $Z$=0.008. Short dashed lines represent a slope of 2.35 and $M_{up} = 30\msun$ for $Z$=0.04 and $Z$=0.008.}
\end{figure*}
\vspace{.2in}

\begin{figure*}
\setcounter{figure}{7}
\epsscale{0.85}
\plotone{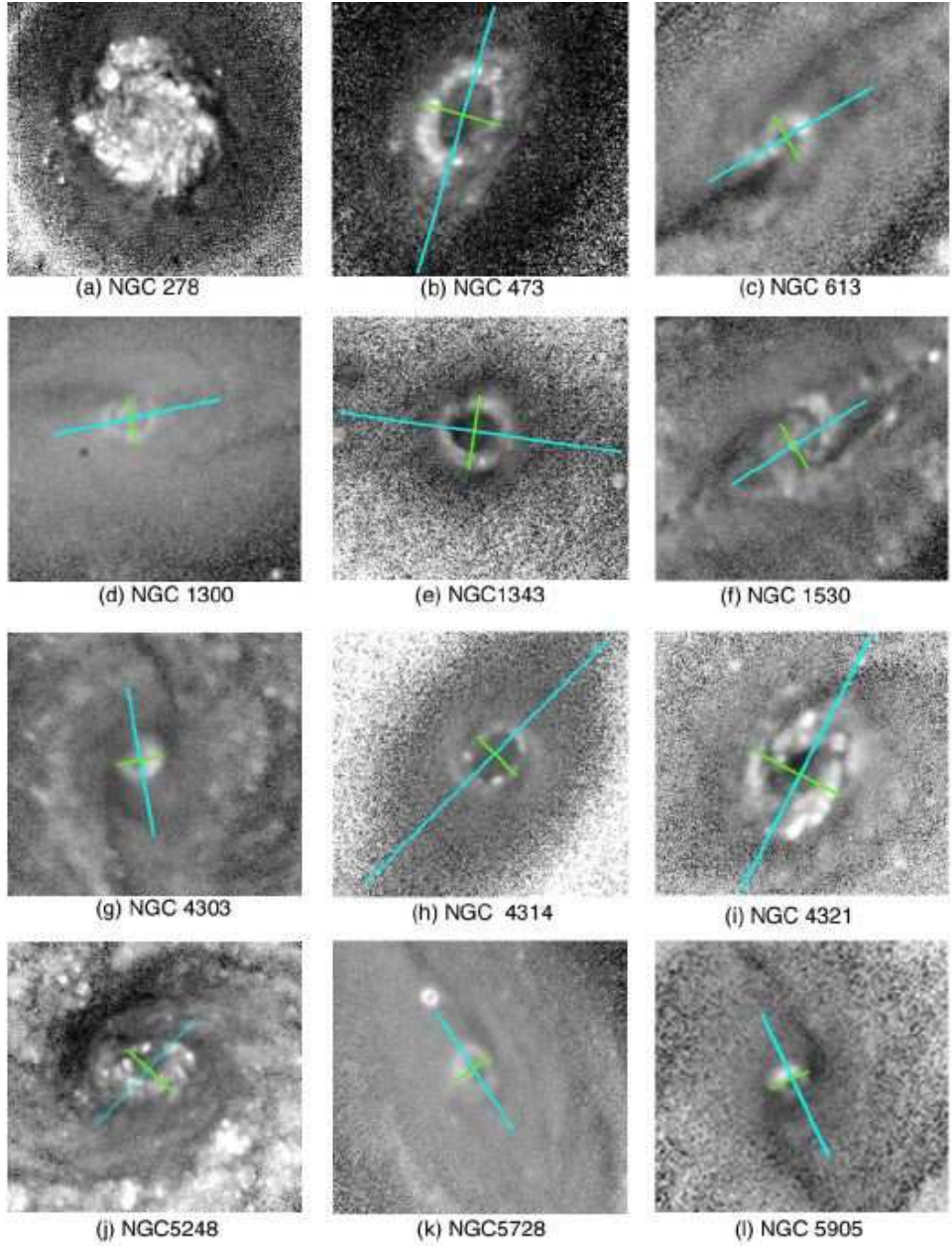}
\caption{panels a. through l.: $B$~-$I$ images with indicators of the bar PA and contact point (CP) PA. The CP PA is defined as a 90$\deg$ offset from the bar PA. In those images with a bar, the blue line (the longer line if viewing in black and white) extends to approximately half the bar major axis size (and traverses the entire FOV in some cases where the ring was magnified such that the bar cannot be fully seen);  the  the green line (shorter line drawn to the approximate ring diameter) represents the CP PAs, which are assumed to be 90 \deg\ offsets of the bar PA.}
\end{figure*}

\begin{figure*}
\setcounter{figure}{7}
\epsscale{1}
\plotone{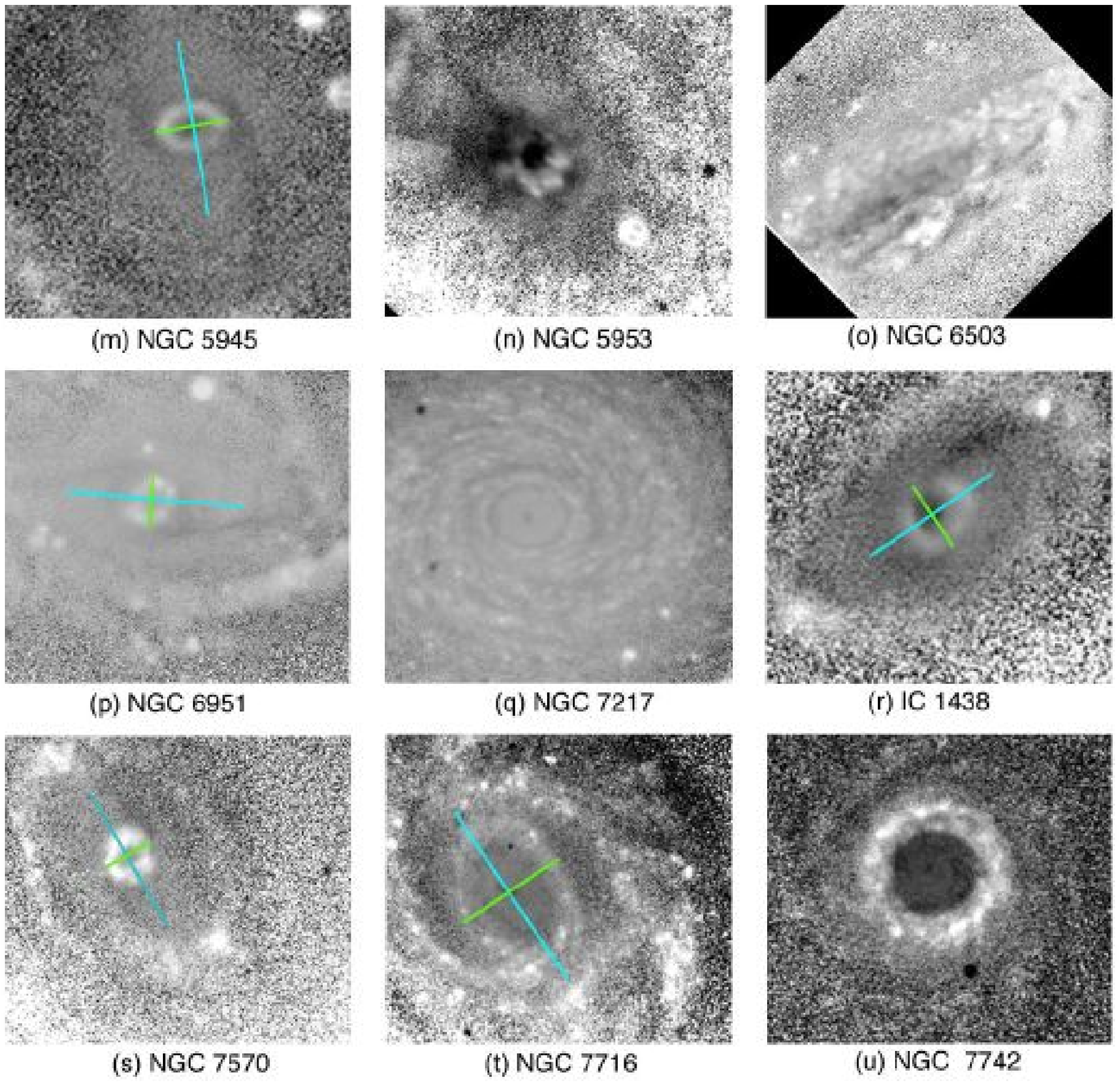}
\end{figure*}

\begin{figure*}
\setcounter{figure}{8}
\plotone{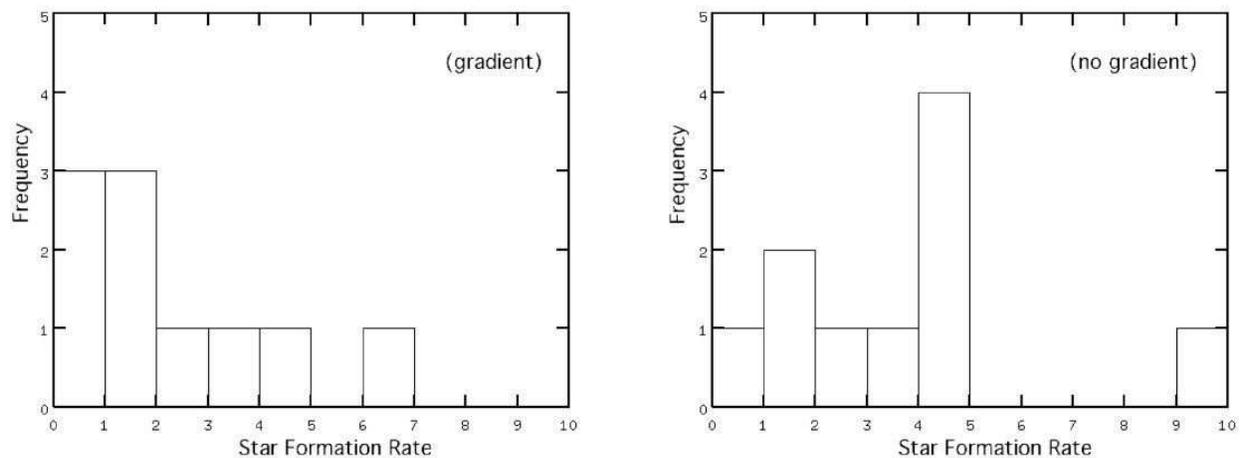}
\caption{Integrated SFR of the rings. The SFR is compared to the presence (left panel) or absence (right panel) of an age gradient. The mean SFR values are 2.2$\pm$0.7 $\msunpyr$ (uncertainty is standard error) for the galaxies with and 3.6$\pm$1.1 $\msunpyr$ for those without gradients.}
\end{figure*}
\vspace{.2in}

\begin{figure*}
\setcounter{figure}{9}
\epsscale{0.55}
\plotone{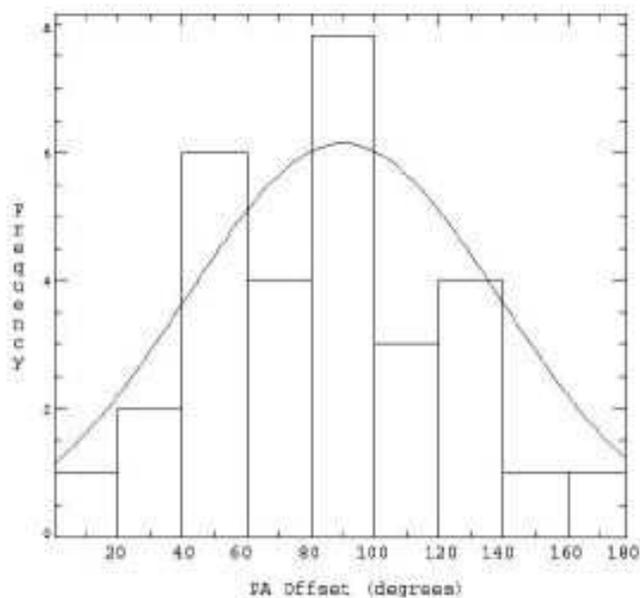}
\caption{Histogram showing the distribution of angular offsets between the bar major axis PA and the locations of the youngest H{\sc ii} region in each hemisphere, which is bisected by the bar. A Gaussian fit to the distribution is overlaid, and shows a peak at an offset of 90\deg, which is consistent with the location of the contact points. See Table~2 for more details.}
\end{figure*}


\clearpage


\begin{thebibliography}{}
\bibitem[Allard et al.(2005)]{2005ApJ...633L..25A} Allard, E.~L., Peletier, R.~F., \& Knapen, J.~H.\ 2005, \apjl, 633, L25 
\bibitem[Allard et al.(2006)]{}Allard, E.~L.,Knapen, J.~H., Peletier, R.~F., \& Sarzi, M. 2006, submitted to MNRAS
\bibitem[Athanassoula(2000)]{2000ASPC..221..243A} Athanassoula, E.\ 2000, Stars, Gas and Dust in Galaxies: Exploring the Links, 221, 243 
\bibitem[Athanassoula(1992)]{1992MNRAS.259..345A} Athanassoula, E.\ 1992, \mnras, 259, 345
\bibitem[Benedict et al.(1993)]{1993AJ....105.1369B} Benedict, G.~F., et al.\ 1993, \aj, 105, 1369 
\bibitem[Bertin \& Arnouts (1996)]{ba96} Bertin, E., Arnouts, S. 1996, \aap, S117, 393
\bibitem[Blackman(1981)]{1981MNRAS.195..451B} Blackman, C.~P.\ 1981, \mnras, 195, 451
\bibitem[Bottema(1989)]{1989A&A...221..236B} Bottema, R.\ 1989, \aap, 221, 236 
\bibitem[Buta \& Combes(1996)]{bc96} Buta, R. \& Combes, F., 1996, Fund. of Cosmic Phys. 17, 95
\bibitem[Buta \& Crocker(1993)]{bc93} Buta, R. \& Crocker, D. A. 1993, AJ, 105, 1344
\bibitem[Combes \& Gerin(1985)]{1985A&A...150..327C} Combes, F., \& Gerin, M.\ 1985, \aap, 150, 327 
\bibitem[Combes et al.(2004)]{2004A&A...414..857C} Combes, F., et al.\ 2004, \aap, 414, 857
\bibitem[Davies \ea (2004)]{dv04} Davies, R. I., Tacconi, L. J., Genzel, R. 2004, ApJ, 602, 148
\bibitem[D\'\i az-Santos et al.]{D06} D\'\i az-Santos, T., Alonso-Herrero, A., Colina, L., Ryder, S.~D., Knapen J.~H. 2007,\apj, 661, 149
\bibitem[Elmegreen(1996)]{1996ASPC...91..197E} Elmegreen, B.\ 1996, IAU  Colloq.~157: Barred Galaxies, 91, 197 
\bibitem[Elmegreen(2002)]{2002ASPC..285..425E} Elmegreen, B.~G.\ 2002, Modes of Star Formation and the Origin of Field Populations, 285, 425
\bibitem[Elmegreen(1997)]{el97} Elmegreen, B. 1997, RMxAC, 6, 165
\bibitem[Garc{\'{\i}}a-Burillo et al.(2005)]{2005A&A...441.1011G} Garc{\'{\i}}a-Burillo, S., Combes, F., Schinnerer, E., Boone, F., \& Hunt, L.~K.\ 2005, \aap, 441, 1011 
\bibitem[Genzel et al.(1995)]{1995ApJ...444..129G} Genzel, R., Weitzel, L., Tacconi-Garman, L.~E., Blietz, M., Cameron, M., Krabbe, A., Lutz, D., \& Sternberg, A.\ 1995, \apj, 444, 129
\bibitem[Helfer et al.(2003)]{2003ApJS..145..259H} Helfer, T.~T., Thornley, M.~D., Regan, M.~W., Wong, T., Sheth, K., Vogel, S.~N., Blitz, L., \& Bock, D.~C.-J.\ 2003, \apjs, 145, 259
\bibitem[Heller \& Shlosman(1994)]{HS94} Heller, C.~H., \& Shlosman, I.\ 1994, \apj, 424, 84 
\bibitem[Heller \& Shlosman(1996)]{HS96} Heller, C.~H., \& Shlosman, I.\ 1996, \apj, 471, 143 
\bibitem[Hummel et al.(1987)]{1987A&A...172...32H} Hummel, E., van der Hulst, J.~M., \& Keel, W.~C.\ 1987, \aap, 172, 32 
\bibitem[Jogee et al.(2002)]{2002ApJ...575..156J} Jogee, S., Shlosman, I., Laine, S., Englmaier, P., Knapen, J.~H., Scoville, N., \& Wilson, C.~D.\ 2002, \apj, 575, 156 
\bibitem[Jogee et al.(2005)]{2005ApJ...630..837J} Jogee, S., Scoville, N., \& Kenney, J.~D.~P.\ 2005, \apj, 630, 837 
\bibitem[Kenney et al.(1992)]{1992ApJ...395L..79K} Kenney, J.~D.~P., Wilson, C.~D., Scoville, N.~Z., Devereux, N.~A., \& Young, J.~S.\ 1992, \apjl, 395, L79 
\bibitem[Kenney et al.(1993)]{1993ApJ...418..687K} Kenney, J.~D.~P., Carlstrom, J.~E., \& Young, J.~S.\ 1993, \apj, 418, 687 
\bibitem[Kennicutt(1998)]{1998ARA&A..36..189K} Kennicutt, R.~C.\ 1998, \araa, 36, 189 
\bibitem[Knapen \ea (1995)]{kkn95} Knapen, J.H., Beckman, J. E., Heller, C. H., Shlosman, I., de Jong, R.S. 1995a, ApJ, 454, 623
\bibitem[Knapen et al.(1995)]{1995ApJ...443L..73K} Knapen, J.~H., Beckman, J.~E., Shlosman, I., Peletier, R.~F., Heller, C.~H., \& de Jong, R.~S.\ 1995b, \apjl, 443, L73 
\bibitem[Knapen(1999)]{1999ASPC..187...72K} Knapen, J.~H.\ 1999, The  Evolution of Galaxies on Cosmological Timescales, 187, 72 
\bibitem[Knapen et al.(1999)]{1999Ap&SS.269..605K} Knapen, J.~H., Laine, S., \& Rela{\~n}o, M.\ 1999, \apss, 269, 605
\bibitem[Knapen et al.(2000)]{2000ApJ...528..219K} Knapen, J.~H., Shlosman, I., Heller, C.~H., Rand, R.~J., Beckman, J.~E., \& Rozas, M.\ 2000, \apj, 528, 219 
\bibitem[Knapen et al.(2004)]{2004A&A...423..481K} Knapen, J.~H., Whyte, L.~F., de Blok, W.~J.~G., \& van der Hulst, J.~M.\ 2004, \aap, 423, 481
\bibitem[Knapen(2005)]{2005A&A...429..141K} Knapen, J.~H.\ 2005, \aap, 429, 141 
\bibitem[Knapen \ea (2006)]{kn06} Knapen,J.H., Mazzuca, L.M., B\"oker, T, Shlosman, I., Colina, L. Combes, F., Axon, D.J. 2006, \aap, 448, 489 
\bibitem[Kormendy \& Kennicutt]{kk04} Kormendy, J., Kennicutt, R. C. Jr. 2004 ARA\&A, 42, 603
\bibitem[Kroupa(2002)]{2002ASPC..285...86K} Kroupa, P.\ 2002, Modes of Star
Formation and the Origin of Field Populations, 285, 86
\bibitem[Leitherer \ea (1999)]{le99} Leitherer, C., et al. 1999, ApJS, 123, 3
\bibitem[Maciejewski et al.(2002)]{2002MNRAS.329..502M} Maciejewski, W.,  Teuben, P.~J., Sparke, L.~S., \& Stone, J.~M.\ 2002, \mnras, 329, 502 
\bibitem[Maoz et al.(2001)]{2001AJ....121.3048M} Maoz, D., Barth, A.~J., Ho, L.~C., Sternberg, A., \& Filippenko, A.~V.\ 2001, \aj, 121, 3048 
\bibitem[Martinet(1995)]{1995FCPh...15..341M} Martinet, L.\ 1995, Fundamentals of Cosmic Physics, 15, 341 
\bibitem[Moiseev et al.(2004)]{2004A&A...421..433M} Moiseev, A.~V., Vald{\'e}s, J.~R., \& Chavushyan, V.~H.\ 2004, \aap, 421, 433 
\bibitem[Oke(1990)]{ok90} Oke,J., 1990 AJ 99,5
\bibitem[Osmer et al.(1974)]{1974ApJ...192..279O} Osmer, P.~S., Smith, M.~G., \& Weedman, D.~W.\ 1974, \apj, 192, 279 
\bibitem[Peterson et al.(1978)]{1978ApJ...226..770P} Peterson, C.~J., Roberts, M.~S., Rubin, V.~C., \& Ford, W.~K.\ 1978, \apj, 226, 770 
\bibitem[Phillips(1996)]{1996ASPC...91...44P} Phillips, A.~C.\ 1996, ASP Conf.~Ser.~ 91: IAU Colloq.~157: Barred Galaxies, 91, 44 
\bibitem[Piner \ea (1995)]{pst95} Piner, B., Stone, J., Teuben, P. 1995, ApJ,449, 508P
\bibitem[Pogge(1989)]{1989ApJ...345..730P} Pogge, R.~W.\ 1989, \apj, 345, 730
\bibitem[Regan et al.(1996)]{1996AJ....112.2549R} Regan, M.~W., Teuben, P.~J., Vogel, S.~N., \& van der Hulst, T.\ 1996, \aj, 112, 2549 
\bibitem[Regan \& Teuben(2004)]{2004ApJ...600..595R} Regan, M.~W., \& Teuben, P.~J.\ 2004, \apj, 600, 595
\bibitem[Regan et al.(1997)]{1997ApJ...482L.143R} Regan, M.~W., Vogel, S.~N., \& Teuben, P.~J.\ 1997, \apjl, 482, L143
\bibitem[Regan \& Teuben (2003)]{re03} Regan, M. W., Teuben, P. 2003 APJ, 582, 723
\bibitem[Rubin(1980)]{1980ApJ...238..808R} Rubin, V.~C.\ 1980, \apj, 238, 808 
\bibitem[Rubin et al.(1997)]{1997AJ....113.1250R} Rubin, V.~C., Kenney, J.~D.~P., \& Young, J.~S.\ 1997, \aj, 113, 1250 
\bibitem[Ryder et al.(2001)]{2001MNRAS.323..663R} Ryder, S.~D., Knapen, J.~H., \& Takamiya, M.\ 2001, \mnras, 323, 663
\bibitem[Sakamoto et al.(1999)]{1999ApJ...525..691S} Sakamoto, K., Okumura, S.~K., Ishizuki, S., \& Scoville, N.~Z.\ 1999, \apj, 525, 691 
\bibitem[Salpeter(1955)]{sal55} Salpeter, E. E. 1955, ApJ, 121, 161
\bibitem[]{}Sarzi, M., Allard, E.L., Knapen, J. H., Mazzuca, L. M., MNRAS in press
\bibitem[Schwarz(1981)]{1981ApJ...247...77S} Schwarz, M.~P.\ 1981, \apj, 247, 77 
\bibitem[Shlosman et al.(1990)]{S90} Shlosman, I., Begelman, M.~C., \& Frank, J.\ 1990, \nat, 345, 679 
\bibitem[Sheth et al.(2000)]{2000ApJ...532..221S} Sheth, K., Regan, M.~W., Vogel, S.~N., \& Teuben, P.~J.\ 2000, \apj, 532, 221 
\bibitem[Smith et al.(1999)]{sm99} Smith, D.~A., Herter, T., Haynes, M.~P., \& Neff, S.~G.\ 1999, \apj, 510, 669 
\bibitem[Storchi-Bergmann et al.(1996)]{1996ApJ...460..252S} Storchi-Bergmann, T., Wilson, A.~S., \& Baldwin, J.~A.\ 1996, \apj, 460, 252  
\bibitem[Tully(1988)] {tl88} Tully, R. B. 1988 British Astronomical Association Journal, 98, 6

\end{thebibliography}
\end{document}